\documentstyle[aaspp4,flushrt]{article}    % preprint
\begin{document}
\slugcomment{ Astrophys.J.Supplement {\bf 111}, 73, 1997}

\title
{ADAPTIVE REFINEMENT TREE -- A NEW HIGH-RESOLUTION $N$-BODY CODE FOR
 COSMOLOGICAL SIMULATIONS}
\author{
 Andrey V. Kravtsov, and Anatoly A. Klypin}
\affil
 {Astronomy Department, New Mexico State University, Box 30001, 
   Department 4500,\\ Las Cruces, NM 88003-0001, USA}

\and

\author {Alexei M. Khokhlov}
\affil {Laboratory for Computational Physics and Fluid Dynamics,
        Code 6404, Naval Research Laboratory,\\ Washington, DC 20375, USA}

%________________________________________________________________________
%

\begin{abstract}
 We present a new high-resolution $N$-body algorithm for cosmological 
simulations. The algorithm employs
a traditional particle-mesh technique on a cubic grid and
successive multilevel relaxations on the finer meshes, introduced 
recursively in a {\em fully adaptive} manner in the regions where the density 
exceeds a predefined threshold. The mesh is generated to  
effectively match an {\em arbitrary} geometry of 
the underlying density field -- a property particularly 
important for cosmological 
simulations. In a simulation the mesh structure is not created
at every time step
but is properly adjusted to the evolving particle distribution. 
 The algorithm is fast and effectively parallel: the 
gravitational relaxation solver is approximately half as fast 
as the fast Fourier transform solver on the same number of mesh cells.
The required CPU time 
scales with the number of cells, $N_c$, as $\sim O(N_c)$. The code 
allows us to 
improve considerably the spatial resolution of the particle-mesh code
without loss in mass resolution. 
We present a detailed description of the methodology, 
implementation, and tests of the code.
 
We further use the code to study the structure
of dark matter halos in high-resolution ($\sim 2h^{-1}$ kpc) 
simulations of standard 
CDM ($\Omega=1$, $h=0.5$, $\sigma_8=0.63$) and
$\Lambda$CDM ($\Omega_{\Lambda}=1-\Omega_0=0.7$, $h=0.7$, $\sigma_8=1.0$)
models. We find that halo density profiles in both CDM and $\Lambda$CDM
models are well fitted by the analytical model presented recently by 
Navarro et al., which predicts a singular 
[$\rho(r)\propto r^{-1}$] behavior of the halo density profiles at small radii.
We therefore conclude that halos formed in the $\Lambda$CDM model
have structure similar to CDM halos and thus
cannot explain the dynamics of the central parts of dwarf spiral galaxies,
as inferred from the galaxies' rotation curves.

\end{abstract}

\keywords{methods: numerical -- cosmology: theory -- dark matter}
\vskip 4cm
%----------------------
\section {Introduction}
%----------------------

$N$-body techniques are used in cosmological simulations 
to follow the nonlinear evolution of a 
system of particles, and to give theoretical
predictions about the matter distribution that can be compared with 
observations. The traditional $N$-body methods are the particle-mesh (PM), 
particle-particle/particle-mesh (P$^3$M), 
and TREE methods (Hockney \& Eastwood 1981; Klypin \& Shandarin 1983;
Efstathiou et al. 1985; Bouchet \& Hernquist 1988,  and references therein). 
Although numerous fundamental results have been obtained using these codes,
the codes often cannot provide
desirable spatial or mass resolution {\em with currently available
computers} because of either memory or CPU limitations. 
Thus, for example, the PM code can handle a large number of particles 
(the latest PM simulations follow evolution of approximately $6\times 10^7$
particles) but is limited in spatial resolution (to increase 
resolution by a factor of 2 requires 8 times as much memory; 
the largest PM simulations
have reached dynamic range of $\sim 1500$). TREE and P$^3$M codes 
are CPU limited\footnote{Although in the case of relatively small number of particles,
$N_p\leq 256^3$, the TREE and P$^3$M codes can provide 
higher dynamic range than the PM code}
because calculation of forces in these codes is considerably 
slower than in the PM code and, in the case of the P$^3$M code, 
is also strongly 
dependent on the degree of particle clustering. 
In an ideal cosmological simulation  one needs 
a resolution $\sim 1-10$ kpc to resolve a galaxy and a simulation cube
of $\sim 100$ Mpc to sample appropriately the longest perturbation waves
or to get sufficient statistics. The number of particles should be 
sufficiently large (usually a few million or larger) to allow halo properties 
to be reliably determined. 
The required dynamical range is thus $\sim 10^4-10^5$, which is higher
than the above codes can provide for the required number of particles and 
with currently available computers. These limitations have 
motivated the development 
of new methods with better resolution and/or performance. 

Villumsen (1989) developed a code in which
the PM grid was complemented by finer cubic subgrids to increase the force 
resolution in regions of interest. The local potential 
was calculated as a sum of the potentials on the subgrids and on the PM
grid. A similar approach was adopted by Jessop, Duncan, \& Chau (1994)
in their particle-multiple-mesh code. However, instead of summing 
the potentials from subgrids, the potential on each level was obtained
independently by solving the boundary problem. 
Boundary values of the potential
were interpolated from the coarser parent grid. 
Couchman (1991) used cubic refinement grids to improve the performance of 
the P$^3$M algorithm. Here, the resolution of the P$^3$M code was
retained while the computational speed was considerably 
increased. In the Lagrangian approach (Gnedin 1995; Pen 1995) 
the computational mesh is not static but moves with the 
matter so that the resolution increases (smaller mesh cells) in the high
density regions and decreases elsewhere. Although potentially powerful,
this approach has its caveats and drawbacks (Gnedin \& Bertschinger 1996). 
The mesh 
distortions, for example, may introduce severe force anisotropies.
A different approach was adopted by Xu (1995), who developed the TPM code,
a hybrid of the PM and TREE algorithms. 
The gravitational forces in the TPM are 
calculated via a PM scheme on the grid and via multipole expansions (TREE 
algorithm) in the regions
where higher force resolution is desired. The forces on the particles
in low-density regions are calculated by the PM scheme, 
while forces on the particles in high-density regions
are {\em the sum} of external large-scale PM force and internal 
short-scale force from the neighboring particles. 
Although this code may not be faster than 
a pure TREE code, it is effectively parallel because particles
in different regions 
can be evolved independently. 
An adaptive multigrid code for cosmological simulations was recently 
presented by Suisalu \& Saar (1995). In this code, finer rectangular 
subgrids are adaptively introduced in regions where the density exceeds a 
specified threshold. For each subgrid, the potential is calculated using
boundary conditions interpolated from the coarser grid. 
The solution on the finer grid is used to improve the 
solution on the coarser grid. Another variant of an adaptive 
particle-multiple-mesh code for cosmological simulations was 
recently presented by Gelato, Chernoff, \& Wasserman (1996). This
code can handle isolated boundary conditions, which makes it applicable 
to noncosmological problems. 

All of the above {\em multigrid} methods use {\em rectangular}
subgrids to increase force resolution. For simulations where 
there are only a few small regions of interest (e.g., a few galaxies 
or clusters of galaxies) the rectangular refinements may be a good choice
because these regions can be easily covered by rectangular subgrids. It is,
however, well known that the geometry of structures in realistic 
cosmological models is usually a complicated network of sheets, 
filaments, and clumps which are difficult to cover efficiently with 
rectangular grids. 

In this paper we present a new Adaptive Refinement Tree (ART)
high-resolution $N$-body code. This code was developed to improve 
the spatial resolution of a particle-mesh code
by about two orders of magnitude without loss of mass resolution or
computational speed. In our scheme, the 
computational volume is covered by a cubic grid that defines
the minimum resolution.
On this grid, the Poisson equation is solved with a traditional fast 
Fourier transform (FFT)
technique using periodic boundary conditions.
 The finer meshes\footnote{We will use the word {\em grid} to refer to
cubic or rectangular configurations, reserving the word {\em mesh} for 
configurations of arbitrary shape.} are 
built as collections of cubic, non-overlapping cells of various sizes 
organized in {\em octal threaded  trees}
in regions where the density 
exceeds a predefined threshold. Any mesh can be subject to further 
refinements; the local refinement process stops when the density criterion
 is satisfied. Once constructed, 
the mesh, rather than being destroyed at each time step, is 
promptly adjusted to the 
evolving particle distribution. 
To solve the Poisson equation on these refinement meshes, we use
a relaxation method with boundary conditions and an initial
solution guess interpolated from the previous coarser mesh. 
Below we present the 
method (\S 2), describe the code (\S 3), and discuss the tests 
(\S 4). We then compare the code 
with other algorithms (\S 4) and finally apply it to a 
real cosmological problem (\S 5). 

%---------------------
\section {Methodology}
%---------------------

\subsection{Adaptive mesh refinement}

Adaptive mesh refinement (AMR) techniques for solving partial differential 
equations (PDEs) have numerous applications in different fields of 
physics, astrophysics,
and engineering in which large dynamic range is important.  
There are two major approaches in the application of these techniques.
In the first approach (e.g., Berger \& Oliger 1984; 
Berger 1986; Berger \& Colella 1989), 
the computational volume is divided into cubic 
elements (cells), while in the second (e.g., L\"ohner \& Baum 1991)
the cells can have an arbitrary shape. Collections of cells
are used as computational 
meshes on which the PDEs are discretized. We will call
meshes composed of cubic cells {\em regular}, calling meshes 
{\em irregular} otherwise. 
The integration of PDEs is simpler on regular meshes, but 
dealing with complicated boundaries may be a difficult  
problem. With irregular meshes one 
can  handle complicated boundaries
much more easily. The price, however,
is more elaborate algorithms, data structures, and associated CPU and memory 
overhead. A particular choice of the mesh structure is a tradeoff between 
these considerations. In astrophysics
there are no complicated boundaries, and a cubic computational
volume is usually used to model a system. In this case, there is no need
for irregular meshes and it is preferable to use meshes made of cubic cells. 

The regular meshes themselves can be organized in different ways. 
The usual practice is to use regular meshes of cubic or rectangular 
shape  (e.g., Berger \& Colella 1989) organized in arrays (grids),
which allows one to 
simplify data structures and to use standard 
PDE solvers. These arrays can be organized 
in a tree (Berger 1986) to form a multigrid hierarchy. The main 
disadvantage of the grids is that one cannot cover 
regions of complicated shape in an efficient way. 
Moreover, the arrays are an inflexible data structure, and 
the whole refinement hierarchy
should be periodically rebuilt, not adjusted, when dealing with unsteady 
solutions. 

In our approach, we use regular meshes but they are handled 
in a completely different way. 
Cells are treated as individual units which are organized in 
{\em refinement trees} (see \S 3.2). 
Each tree has a root -- a cell belonging to a base cubic grid that
covers the entire computational volume. 
If the root is refined (split) -- it has eight 
children (smaller nonoverlapping cubic cells residing in its volume), 
which can be refined in their turn, and so on. Cells of a given 
{\em refinement level} are organized in {\em linked lists} 
and form a refinement mesh. 
The tree data structures make mesh storage and access in memory logical
and simple, while linked lists allow for efficient mesh structure 
traversals. In the current version of the code we make use of 
octal threaded trees (Khokhlov 1997) and doubly linked lists 
(e.g., Knuth 1968; Aho, Hopcroft, \& Ulman 1983; 
Corner, Leiserson, \& Rivest 1994). 
The fact that cells are treated as independent units rather than 
element of a grid allows us to build a very flexible
mesh hierarchy which can be easily modified. The details of the mesh
generation and modification in our code are
described in \S3. 
\subsection{Multilevel relaxation method}
The multigrid techniques of solving partial 
differential equations (Brandt 1977)
are very successful in reducing the 
computational and storage requirements for solving many types of PDEs 
(Wesseling 1992; Press et al. 1992). There are two kinds of
multigrid algorithms. The first, 
sometimes called {\em the multigrid method},
is used to speed up convergence of relaxation methods. 
In this method, the source term is defined only on 
the base finest grid -- all the other, coarser grids are used as a workspace. 
In the second algorithm, called {\em full multigrid}, the source term is 
defined on all grids, and the method obtains successive 
solutions on finer and finer grids. The latter method is 
useful when dealing with grids created in adaptive refinement process.
The full multigrid scheme can be used differently in its turn, depending on 
how the solutions on different levels influence each other. In 
the {\em one-way interface} scheme, the solution from 
a coarser 
grid is used to get a first-guess solution on the finer grid and often 
to get boundary values as well. However, the solution on the coarser grid 
is not influenced by the solution on the finer grid 
(e.g., Jessop, Duncan, \& Chau 1994). In the {\em two-way} 
interface scheme, the coarser grid solution is used to 
correct the solution on the finer 
grids and vice-versa. The choice of a particular scheme is usually determined 
empirically and is problem dependent. The two-way interface scheme
is more difficult to implement in the case of periodic boundary 
conditions (Suisalu \& Saar 1997). 
 
In our approach, each refinement mesh is composed
of cells of the same refinement level, but these meshes are completely 
different from grids. The techniques are thus 
multilevel rather than multigrid. 
We use an analog of the full multigrid algorithm with 
a one-way interface between the 
meshes. We use a regular cubic grid covering the whole computational volume 
as the zeroth or coarsest level. At this level, the Poisson equation is solved
 using a standard FFT  method with periodic boundary conditions. 
This solution is then interpolated onto the  
finer first-level mesh to get the boundary values
and first-guess solution. Once the boundary problem is defined, we use
a {\em relaxation} method (e.g., Press et al. 1992) to solve the 
Poisson equation on the mesh. Since we start from an initial guess 
which is already close to the final solution, the 
iterative relaxation procedure converges quickly. 
After we get the solution on the first refinement level, the same
procedure (obtaining boundary values and initial guess by interpolation
from the previous coarser level) is repeated for
the next level, and so forth. 
At the end of this process we have the solution (potential)
for all cells. The description of the code is given in the next section.

%--------------------------------
\section {Description of the code}
%--------------------------------
\subsection{Code structure}
The structure of the code can be outlined as follows.
First of all, we set up the initial positions and velocities of the particles 
using the Zeldovich approximation, as described by Klypin \& Shandarin (1983). 
Once the initial conditions are set, we construct the regular cubic grid 
covering the whole computational volume and then proceed to check whether
additional refinement levels are required according to the current density 
threshold. At this point the code enters the main computational loop, which 
includes:
\begin{itemize}
\item density assignment on all existing meshes;
\item a gravitational solver;
\item routine updating of particle positions and velocities;
\item modifications to the mesh hierarchy.
\end{itemize}
The mesh modifications (refinement and derefinement) are based on the
density distribution\footnote{The density criterion is a natural
choice in our case  because we aim to resolve high-density regions.   
We could use, however, any other appropriate criterion, e.g., local potential 
gradient, force accuracy, etc.}. 
The modifications are made at the end of the computational cycle.
At this point the density distribution is 
available, since it was calculated for the gravitational solver.

Below we will describe each of these major functional blocks in detail.
We will also discuss timing, 
energy conservation, and the memory requirements of the code.

\subsection{Mesh generator.}
The adaptive mesh refinement block of the code generates new meshes
and modifies existing ones. The refinement hierarchy 
in our implementation
is based on the regular cubic grid that covers the entire computational 
volume. With the refinement
block turned off, the density assignment
and gravity solver on this grid are similar to those in the PM code 
of Kates, Kotok, \& Klypin (1991). 

The data structures that we use to organize the mesh cells 
are very similar to those implemented in the hydrodynamical 
Eulerian tree refinement code\footnote{There are, however, some important 
modifications required by specifics of the cosmological 
simulations.} (Khokhlov 1997).
All mesh cells are organized in {\em refinement trees}.
A cell can be a {\em parent} of eight {\em children}
 -- smaller cubic cells of equal volume residing in it. Each child may
be in its turn split and have children. 
 Each tree has 
a {\em root} (a zeroth-level cell) that may be the only cell in this tree 
if it is unsplit. The tree ends with unsplit cells, which we call 
{\em leaves}. This structure is called 
an {\em octal rooted tree}, 
and is the construct used in TREE codes.
There is, however, an important difference between our code
and TREE codes. We use 
{\em fully threaded trees},
in which cells are connected with each other {\em on all 
levels}.  
In addition, cells that belong to different trees are 
connected to each other across tree boundaries.
In fact, we can consider all cells as belonging to a single threaded 
tree with a root being the entire computational domain and the base 
grid being one of the tree levels. 
 The tree structure
is supported through a set of pointers. Each cell has a 
pointer to its
parent and a pointer to its first child. In addition, cells have pointers to 
the six adjacent cells (these make the tree fully threaded) 
so that information about a cell's neighbors is easily accessible (see Fig.1). 
Overall, the following information is 
provided for each cell $i$ belonging to a tree:
\begin{itemize}
 \item $Level(i)$, the level of the cell in the tree; 
 \item $Parent(i)$, the pointer to the parent cell;    
 \item $Child(i)$, the pointer to the cell's first child,  
                        or $nil$ if the cell is a leaf;
 \item $Nb(i,j)$, pointers to neighboring cells
                        ($j=1,...,6$);
 \item $Pos(i,j)$, position in space ($j=1,...,3$);
 \item $Var(i,n)$, storage for associated physical 
                        variables (in our case $n=2$, as we store
                        both the density and the potential).  
\end{itemize} 
The above set of pointers is sufficient to support the tree structure 
and to change it dynamically with minimum cost. In addition, the cells on
each level of the mesh hierarchy are organized in doubly linked 
lists\footnote{The difference between a doubly linked list and the usual 
linked list (used, for example, in the P$^3$M codes) is that in the former
we keep not only a pointer to the next element but also a pointer to the 
previous element in the list. This allows us to insert and delete 
list entries without rebuilding the whole list.}
(e.g., Knuth 1968) so that 
a sweep through a given level (the operation used extensively in the multigrid 
relaxations described below) can be done with minimum CPU time.
This organization adds two pointers for each eight 
siblings, or $1/4$ storage elements per cell. 
The cells belonging to the base regular grid (level zero), 
while part of the same data structure as the other cells, 
are created only at the very beginning of a simulation and are
never destroyed.
It is therefore unnecessary to keep information about a cell's 
position or pointers
to its neighbors because they can be easily computed. 
The number of pointers can be considerably 
 reduced (by as much as a factor of 2) 
because some of them can be shared by siblings 
(sets of eight cells with the same parent). 

An elementary refinement process creates eight new cubic cells of equal 
volume ({\em children}) inside a {\em parent} cell. When the parent
is refined, we check if all six neighbors are of the same level as
the parent. If there are coarser neighbors (of smaller level than 
the parent), we split those neighbors. If a neighbor in its turn has coarser 
neighbors, we split the neighbor's neighbors, and so forth. 
We thus build a refinement structure which obeys a rule allowing 
{\em no neighbor cells with level difference greater than} 1.
Examples of allowed and prohibited configurations are shown in Figure 2.
Although this is the only rule in the whole refinement process,
it determines the global structure of the resulting refinement hierarchy,
assuring that there are no sharp resolution gradients on a level's boundaries.
On the next refinement pass, each of the newborn children is checked 
against the density criterion and can be subdivided into 
8 children  in its turn if further splitting is needed. 
The process stops when either the density criterion
is satisfied everywhere or the maximum allowed refinement level is reached. 

The refinement process proceeds level by level starting from 
the base grid. On any level of the mesh hierarchy the process can 
be split into two major parts. First, we mark up\footnote{In this step,
a cell is marked for splitting if the local density exceeds a predefined 
level-dependent threshold.} all the cells which need to 
be split, creating a refinement map. However,
a map constructed in this way tends to be ``noisy''. We smooth it by 
marking additional cells so that any cell which was  originally marked is  
surrounded by a buffer of at least two other marked cells.
We construct this buffer using an algorithm
which includes several passes through a level, each one
marking additional cells. 
During the first pass the neighbors of cells marked in 
the refinement map are marked for splitting also. 
After that, two passes are made in which we mark for splitting only those cells
which have at least two neighbors already marked for refinement (note 
that when we speak of marked cells, we mean cells marked only during passes
before the one we are discussing, not during the pass under
consideration). 
These three passes create a one-cell cubic buffer around each of the
cells marked in the original refinement map. Each additional set of three
passes similar to those described above will build one more cubic 
layer around every originally marked cell. 
Therefore, to build a two-cell buffer we make six passes. 
When the map is completed, it is used to make the actual splitting. 

The refinement procedure described above can be used either to construct the 
mesh hierarchy from scratch or to modify the existing meshes. However,
in the course of a simulation the structure is neither constructed 
nor destroyed. Instead, in every computational cycle we {\em modify} existing
meshes to account for the changes in particle distribution. Therefore,
we need to make not only refinements but also derefinements (in
the places where it is no longer necessary to keep resolution at the
current level), which is accomplished in the same manner as refinement 
by constructing a 
derefinement map -- that is, a map of cells marked for joining. 
If the joining violates the above-mentioned neighbor rule, nothing is done 
and the cell remains split. Therefore, the code modifies the existing 
structure dynamically, keeping the refinements in accord with the ever-changing
density field. Modifying the hierarchy requires much less
CPU time than rebuilding it because only a small number of cells
needs to be modified at any given time step. Figure 3 shows an example 
of the refinement mesh hierarchy built in one of the $\Lambda$CDM 
cosmological simulations described in \S 4.4. A selected region of
Figure 3 is shown expanded in Figure 4. 

\subsection{Particles within the mesh hierarchy and density assignment}

Particle coordinates are not sufficient to specify the particle-mesh 
connection because cells of different levels can share the same
volumes; we need to know, however, which particles belong 
to a given cell. We keep track of the particles by arranging them
in doubly linked lists so that 
every cell ``knows'' its head linked-list particle (the head is nil if 
the cell is empty) and thus all the other particles in this linked list. If 
a particle moves from cell to cell, it is deleted from the linked list of 
the cell it leaves and is added to the new cell's  linked list. 
Only leaves are allowed to own particles. Once a cell is split, all its
particles are divided among its children. 
However, we solve the Poisson equation on every refinement 
level, so that the value of the 
density must be computed for every cell regardless of whether 
or not it is a leaf.
 On each level, starting from the finest level and up to the zeroth level, 
the density is assigned using the standard cloud-in-cell (CIC) technique 
(Hockney \& Eastwood 1981). 
Because particles belong only to the finest cells enclosing
them, when we change between levels the particles are passed from children
to their parents. The particles are transferred in this way only as far
as the density assignment is concerned; the linked list is not
changed. The particles, therefore, contribute to the density on any
level in which they are physically located. 

\subsection{Poisson solver}

The fact that the zeroth level of the mesh hierarchy is a cubic regular grid
of fixed resolution allows us to use the FFT method
to solve the Poisson equation on this grid (Hockney \& Eastwood 1981). 
The FFT technique naturally supports periodic boundary conditions which 
is important for cosmological simulations. Moreover, the FFT is well 
benchmarked and is about twice as fast as the relaxation method described 
below. 

The Poisson equation on the refinement meshes is defined as a 
Dirichlet boundary 
problem for which boundary values are obtained by interpolating the potential 
from the parent grid. In our algorithm, 
the boundaries of the refinement meshes can have an arbitrary shape,
which narrows the range of PDE solvers  
one can use. To solve the Poisson equation on these meshes, we have
chosen the {\em relaxation} method (Hockney \& Eastwood 1981; 
Press et al. 1992), which is relatively fast and efficient in dealing 
with complicated boundaries.
In this method the Poisson equation 
\begin{equation}
\nabla^2 \phi = \rho
\end{equation}
is rewritten in the form of a diffusion equation,
\begin{equation}
\frac{\partial \phi}{\partial \tau}=\nabla^2 \phi - \rho.
\end{equation}
The point of the method is that an initial solution guess $\phi$
{\em relaxes} to an equilibrium solution (i.e., solution of the Poisson 
equation) as $\tau \to \infty$. 
The finite-difference form of equation (2) is:
\begin{equation}
\phi^{n+1}_{i,j,k}=\phi^n_{i,j,k}+\frac{\Delta \tau}{\Delta^2}
\left(\sum^6_{nb=1}\phi^n_{nb}-6\phi^n_{i,j,k}\right)
-\rho_{i,j,k}\Delta \tau.
\end{equation}
where the summation is performed over a cell's neighbors.
Here, $\Delta$ is the actual spatial resolution of the solution 
(potential), while $\Delta \tau$ is a fictitious time step 
(not related to the actual time integration of the $N$-body system). 
This finite difference method is stable when
$\Delta \tau\leq\Delta^2/6$ (Press et al. 1992). 
If we choose the maximum allowed time step $\Delta \tau=\Delta^2/6$, 
the above equation can be rewritten in the form of the following 
iteration formula:
\begin{equation}
\phi^{n+1}_{i,j,k}=\frac{1}{6}
\left(\sum^6_{nb=1}\phi^n_{nb}-6\phi^n_{i,j,k}\right)
-\frac{\Delta^2}{6}\rho_{i,j,k}.
\end{equation}
The relaxation iteration thus averages the potential 
of a cell's  six neighbors and subtracts the contribution from the source
term. Cells in the boundary layer will have some neighbors
belonging to the coarser level. In this case, we 
need to interpolate to get the potential at the location of the 
expected neighbor. It is desirable that the interpolation maintain 
continuity and isotropy of the force (see discussion in Jessop et al. 1994). 
We have found that linear interpolation perpendicular
to the boundary which incorporates both coarser and finer cell potentials
is satisfactory; we get the interpolated value 
of the potential on the boundary of level $l$ as:
\begin{equation}
   \phi_{int}=w_i\phi_l+(1-w_i)\phi_{l-1}.
\end{equation}
Here $w_i$ is a weight, and $\phi_l$ and $\phi_{l-1}$ 
are the potentials of a boundary cell 
of level $l$ and of its $(l-1)$-level neighbor.
We found the optimal value of $w_i$ to be $0.2$ by minimizing the force
discontinuity for particles moving through mesh boundaries. 
The iterative procedure described above is repeated until 
the desired level of convergence is achieved. 
We can speed up the convergence of the 
relaxation procedure considerably by using an initial guess for the 
solution that is already 
close to the final solution. Such an initial guess can be 
obtained by interpolating the potential from the previous coarser mesh,
for which the Poisson equation was already solved. By doing so, we 
need only $2-3$ iterations to find the potential to an accuracy 
of a one or two percent. Nevertheless, a higher accuracy is needed because
the potential is then differentiated to get the accelerations; 
the errors in accelerations are thus larger than the errors in 
the potential. Therefore, we would need to make more iterations to reach 
the same $\sim 1-2\%$ accuracy level in the acceleration. 
The number of required iterations, however, can be considerably 
reduced by using the so-called {\em successive overrelaxation} (SOR)
technique (Hockney \& Eastwood 1981; Press et al. 1992). 
In this technique, the solution 
in a given cell is computed as a weighted average,
\begin{equation}
\phi^{n+1}_{\ast} = \omega \phi^{n+1} + (1-\omega)\phi^n,
\end{equation}
where $\phi^{n+1}$ is the solution obtained via the iteration equation (4),
$\phi^n$ is the solution from previous iteration step, 
and $\omega$ is the {\em overrelaxation parameter}. 
The parameter $\omega$ can be adjusted to minimize the number of 
iterations required to achieve a certain accuracy level. 

The ultimate goal of any $N$-body algorithm is to get an accurate
approximation to the pairwise interparticle forces.  Therefore, we
use the force accuracy (see \S4) to determine the required 
number of iterations. 
Of course, there is no point in using more iterations than the number 
needed to make
the iteration error smaller than truncation error. The latter can 
be estimated by making the number of iterations very large, so that 
the iteration error is negligible.
We then find the minimum number of iterations for which 
the force accuracy is still at the level of truncation errors. The number of 
iterations is further minimized by adjusting the overrelaxation parameter.
We have found empirically that only 10 relaxation iterations are needed if 
$\omega =1.25$.

\subsection{Particle dynamics}

To integrate the trajectories of the dark matter particles we use the 
Newtonian equations of motion in an expanding cosmological framework
(e.g., Peebles 1980). These equations can be expressed in terms 
of comoving coordinates ${\bf x}$ related to the 
proper coordinates as ${\bf r}=a(t){\bf x}$, where 
$a(t)=(1+z)^{-1}$ is the expansion factor: 
\begin{equation}
\frac{d{\bf p}}{dt}=-\nabla_x\phi, \ \ \ 
\frac{d{\bf x}}{dt}=\frac{{\bf p}}{a^2},
\end{equation}
where ${\bf p}$ is the  momentum of a particle and
$\nabla_x\phi$ is given by the Poisson equation relating 
the potential $\phi$ to deviations of density from the background:
\begin{equation}
\nabla^2_x\phi=4\pi Ga^2(\rho - \overline{\rho}). 
\end{equation}
The above equations are integrated numerically using dimensionless
variables
\begin{equation}
         {\bf x}=x_0{\widetilde{\bf x}},\ 
         t={\widetilde{t}}/H_0,\ 
         \phi={\widetilde{\phi}}(x_0H_0)^2,\ 
         {\bf p}={\widetilde{\bf p}}(x_0H_0),\ 
         \rho={\widetilde{\rho}}\frac{3H_0^2}{8\pi G}
         \frac{\Omega_M}{a^3},\ \  
\end{equation}
where $x_0$ is the length of a zeroth-level mesh cell and $H_0$ is 
the Hubble constant. We also use the expansion factor $a$ instead of 
the time $t$, so that equations (7)--(8) can be rewritten as:
\begin{eqnarray}
\frac{d{\widetilde{\bf p}}}{da}&=& -f(\Omega_M,\Omega_{\Lambda},a){\ }
{\widetilde{\nabla}}{\mathaccent "7E \phi},\nonumber\\
\frac{d{\widetilde{\bf x}}}{da}&=&f(\Omega_M,\Omega_{\Lambda},a){\ }
\frac{{\widetilde{\bf p}}}{a^2},\\
{\widetilde{\nabla}}^2{\widetilde{\phi}}&=&
\frac{3\Omega_M}{2a}\left({\widetilde{\rho}}-1\right). \nonumber
\end{eqnarray}
Here $\Omega_M$ is the present-day ($z=0$) contribution of matter to the 
total density of the universe and $\Omega_{\Lambda}$ is the corresponding 
contribution of the vacuum energy (measured by the cosmological constant). 
The function $f$ is  specific to a given cosmological model. 
The general form of this function, valid for open, flat, and closed 
cosmologies, is (e.g., Carrol et al. 1992):
\begin{equation}
f(\Omega_M,\Omega_{\Lambda},a)=\frac{1}
{\sqrt{1+\Omega_M\left(1/a-1\right)+\Omega_{\Lambda}(a^2-1)}}.
\end{equation}

We adopt
a standard second-order leapfrog integration scheme of advancing particles
to the next time step. For a  step $n$, corresponding to time step
$a_n=a_{init}+n\Delta a$, the momenta and positions of particles 
are updated as follows:
\begin{eqnarray}
     {\widetilde{\bf p}}_{n+\frac{1}{2}} & = & 
     {\widetilde{\bf p}}_{n-\frac{1}{2}}-
     f(\Omega_M,\Omega_{\Lambda},a_n){\ }{\widetilde{\nabla}}
     {\widetilde{\phi}}_n{\ }\Delta a, \nonumber \\
       &  & \\
     {\widetilde{\bf x}}_{n+1} & = & 
     {\widetilde{\bf x}}_n+
     f(\Omega_M,\Omega_{\Lambda},a_{n+\frac{1}{2}}){\ }
     \frac{{\widetilde{\bf v}}_{n+\frac{1}{2}}}
          {a^2_{n+\frac{1}{2}}}{\ }\Delta a. \nonumber 
  \end{eqnarray}
Here the indices $n$, $n+1$, and $n\pm\frac{1}{2}$ refer to 
quantities evaluated at $a_n$, $a_{n+1}$, and $a_n\pm\Delta a/2$ 
respectively. 
Although multiple time stepping is probably 
very efficient in terms of CPU time, in the current version of the code
we use a constant time step for all particles. We plan to implement 
individual time steps for different levels in the future. 
Particle coordinates and velocities are updated 
using their accelerations obtained via numerical differentiation 
of the potential and interpolation to the particle 
location using the CIC method (Hockney \& Eastwood 1981). 
There are, however, some complications
because particles can move through the level 
boundaries. The resolution gradients, for example, 
may induce unwanted force fluctuations and anisotropies (Jessop et al. 1994;
Anninos, Norman, \& Clarke 1994). In addition, momentum conservation,
achieved by exact cancellation of numerical terms in the CIC method, is 
no longer guaranteed. This means that additional care must be taken 
to minimize these effects. Usually, this is done by introducing 
extra {\em buffer regions} along the mesh interfaces so that 
force interpolation on the boundaries is avoided. In our code, we do not 
introduce additional buffer cells on the mesh boundaries because 
the meshes are already expanded by smoothing (see \S 3.2). 
Therefore, we simply prohibit force interpolation that uses
both coarse and fine boundary cells, interpolating instead on the 
coarse level. In this way, particles are driven by the coarse force
until they move sufficiently far into the finer mesh. The same 
is true for particles moving from the finer to coarser mesh. 

\subsection{Memory requirements}
The memory requirements of the code are determined by the number of 
dark matter particles $N_p$, the number of cells in the zeroth-level base
grid $N_c^0$, and the number of cells on refinement levels $N_c^L$. 
In the current implementation of the code the total number of 
memory storage elements $N$ used by the code is:
\begin{equation}
N^{ART}\approx8N_p+6N_c^0+15N_c^L.
\end{equation}
The particle information consists of coordinates, momenta, and 
two pointers used to organize the doubly linked list. 
The overhead for $N_c^L$ is determined by the pointers used
to support the tree refinement hierarchy (see \S 3.2). It can be 
significantly reduced (this scheme was implemented in Khokhlov 1997)
if most of the cell information (namely $Level$, $Parent$, $Nb$, and $Pos$; 
see \S 3.2) is shared between siblings (the eight cells 
which have the same parent). 
The cells have individual pointers to their 
first child and to two physical variables (density and potential). 
In this case, for eight refinement cells we need only 
$1$Level$+1$Parent$+6$Nb$+3$Pos$+8$Child
$+16$Var$=35$, storage elements ($\approx 4.5$ per cell instead of 15). 
We plan to implement these improvements in the future versions of the code.

$N^{ART}$ can be compared to the corresponding number of 
storage elements in a PM code:
\begin{equation}
  N^{PM}\approx 6N_p+N_c^0;
\end{equation}
The apparent overhead of the ART code compared to the PM code is the 
price for a fully adaptive and flexible mesh structure. 
It should be noted, however, that to increase resolution by a 
factor of 2 in a PM code the number of cells $N_c^0$ must be 
increased by a factor of 8, which severely limits the maximum 
possible dynamic range 
($\sim 1500$, with largest currently available computers). 
In the ART code, the resolution is improved by 
increasing $N_c^L$ which changes very slowly when 
resolution is increased. For example, to increase resolution in 
the highest density regions by a factor of 2 in the simulations 
described in Section 5 (see also Table 1), 
the total number of cells was increased only by $\sim3\%$. 
Note also that the dynamic range of $\sim 4000$ was achieved with only 
$\sim 5\times 10^6$ cells while a PM code would require 
$\sim 6.4\times 10^{10}$ cells to reach the same resolution. 
For comparison, the memory requirements of the publicly available version
of TREE code (kindly provided by J.Barnes) is 
$N^{TREE}\approx 11N_p+18N_{cells},$
where $N_{cells}$ is the number of tree cells. The memory requirements of 
AP$^3$M code (Couchman 1991) are  
$N^{AP^3M}\approx 10N_p$.
%-------------------
\subsection{Timing}
%-------------------
In this section we present timings of the current version of the 
code and compare the performance with other high-resolution
$N$-body codes. 
The present version of the code was parallelized to run in shared memory 
mode on the HP-Convex SPP-1200 Exemplar -- a multipurpose scalable parallel 
computer. Currently, the code is not fully parallelized. The 
blocks of code 
that require considerable parallelization efforts, namely
the density assignment and cell splitting/joining during the mesh 
modifications, are run serially. The parallelization of the most CPU-expensive 
parts of the code, the Poisson solver and the force interpolation,
was straightforward.
We are now working on the complete parallelization (including distributed 
memory architectures) and optimization of the code and will present details 
elsewhere. 

In Figure 5a we show the performance of different blocks 
of the code with respect to the expansion parameter in a 
$\Lambda$CDM simulation
($\Omega_{\Lambda}=1-\Omega_0=0.7$, $h=0.7$, $\sigma_8=1.0$)
of an $L=15h^{-1}$ Mpc box with $N=32^3$ particles and base grid of
$64^3$ cells
(more details are given in \S 4.4). The simulation was run
on 8 CPUs of the NCSA SP-1200 Exemplar in shared memory mode. The CPU overhead
for running code in parallel is about $50\%$ and the same simulation
run on an IBM RS/6000 workstation performs $\sim2.5$ times faster (in terms
of CPU but not in terms of wall-clock time!). The overhead is 
mostly due to the unusually large penalty for cache-missing events that
results if memory is accessed randomly. 
In Table 1 we present timing for different
code blocks for the final time step ($z=0$) of two similar $\Lambda$CDM 
simulations with $64^3$ particles (see section 5.2) and with different 
resolutions. The base grid in both simulations was $128^3$ and 
numbers of maximum  allowed refinement levels were 4 and 6 (the number of 
mesh cells in Table 1 includes zeroth-level cells). 
As before, the simulations were run 
on 8 CPUs of the SPP-1200 Exemplar. 
We compare the timings from Table 1 with the performance of the 
AP$^3$M code (Couchman 1991).
The final step in a two-level AP$^3$M simulation of an 
open ($\Omega_0=0.5$, 
$\Omega_{\Lambda}= 0$, $h=0.63$, $\sigma_8=1.2$) cosmology (box 
$L=150h^{-1}$ Mpc, $128^3$ grid, $128^3$ particles, and smoothing kernel
$\eta =0.1$ cell giving a dynamic range of about $1000$) took 1316 CPU seconds 
on one processor of IBM SP-2 computer (S.Borgani 1996, private communication).
This number is roughly consistent 
with the timings presented in original paper of Couchman (1991) if we account 
for the difference in Mflops between the machines used. The first 
simulation in Table 1 is comparable in spatial resolution to the above 
AP$^3$M simulation but we must account for the 
different number of particles. 
Only density assignment and particle motion directly scale with number of 
particles. We, therefore, multiply the CPU time spent by 
these routines by 8 which gives
a total time for the final step $\sim 650$ CPU seconds. Note, however, that 
about half of this CPU time is penalty for the cache missings which would
be negligible for a serial run on the SP-2. The ART code, therefore, is about
3 times faster than AP$^3$M code of comparable resolution. Note also 
that although in the second simulation from Table 1 the resolution 
was increased
by a factor of 4, the CPU time did not change significantly because 
only a relatively small number of additional cells was required to 
resolve the highest density regions.  
\subsection{Energy conservation}
In an expanding universe, energy conservation is expressed 
by the Irvine-Layzer-Dmitriev-Zeldovich equation,
\begin{equation}
\frac{d}{dt}\left[a(T+U)\right]=-T\frac{da}{dt},
\end{equation} 
or
\begin{equation}
a(T+U)\vert^a_{a_0}=-\int^a_{a_0}Tda,
\end{equation}
where 
\begin{equation}
T=\frac{1}{2}\sum^{N_p}_{i=1}\frac{p_i^2}{a^2},
\ \  \  U=\frac{1}{2}\sum^{N_p}_{i=1}\phi_i.
\end{equation}
The error in energy conservation at a time $a_i$ is then measured by 
comparing the change in total energy with the change in $aU$:
\begin{equation}
Error = \frac{a(T+U)\vert^{a_{i-1}}_{a_0}+\int^{a_{i-1}}_{a_{i-2}}Tda}
{aU\vert^{a_{i-1}}_{a_0}}
\end{equation}
In Figure 5b we show the energy conservation error versus the expansion 
parameter $a$ for two $\Lambda$CDM simulations with $32^3$ and $64^3$ 
particles. In both simulations the time step, $\Delta a$,  was chosen 
so that none of the particles would move more than a fraction 
($\sim0.1-0.3$) of the mesh cell they were residing in over one step. 
We note that energy is conserved at the level of $\sim 2\%$ in the $32^3$ 
particle simulation
and at the level of $\sim 1\%$ in the $64^3$ particle simulation. 
The maximum errors of 
$\sim5\%$ and $\sim 3\%$ for the two simulations occurred when the first 
two refinement levels were opened at $a\sim 0.18$. This may be a result 
of the rather fast change in resolution in the regions of ongoing 
nonlinear collapse. 

%-----------------------------
\subsection{Density threshold}
%-----------------------------

One of the important parameters of the code is the 
density threshold, which is used
to decide whether a given patch of the computational volume should be refined 
or derefined. The usual practice in the adaptive refinement algorithms is to 
consider this threshold a free parameter (see, e.g., Suisalu \& Saar 1995).
While we also take it as a free parameter, we will discuss here some practical 
limits. First of all, we do not want to set up the density threshold too low
because it usually leads to shot noise in the refinement procedure and can 
also lead to unwanted two-body effects (we should not force a resolution 
considerably less than mean separation between particles). 
Our tests (particularly the 
spherical infall test presented in \S 4) have shown that two-body effects 
are negligible and that the refinement/derefinement 
procedure is stable if we use a density 
threshold corresponding to  $\geq 5-6$ particles in a mesh cell. We choose the 
minimum value of $5$ for the simulations described in \S 5. 
We can also consider this parameter from a different point of view. 
The threshold can be thought of as a parameter defining the overdensity at 
which the refinement takes place. In the case of simulations presented in 
\S 5 the threshold value 
corresponds to an overdensity of $40$ on the first refinement 
level. 
This overdensity is reasonable for the purpose of resolving a halo -- 
the first refinement happens well beyond the halo virial radius (corresponding 
to the overdensity $200$). We have also found that results are not particularly
sensitive to the threshold in the range of $5-10$ particles per cell.

%---------------------------
\section {Tests of the code}
%---------------------------

In this section we present tests of the developed code. 
In particular, we show results describing 
the accuracy of the force calculation in ART scheme and related
issues of resolution. We discuss the results of the Zeldovich 
pancake test and of the spherical infall test.  
Finally, we compare results for a set of  realistic 
cosmological runs obtained with the ART and PM codes. 

\subsection{Force accuracy} 

It is important to know the shape
and accuracy of both short- and long-range forces in order to compare 
the resolution of different codes. Here we present a test
showing the accuracy of force calculations in PM and ART 
schemes and compare them to the Plummer softened force 
often used in both P$^3$M and TREE codes.

We used a $64^3$ base grid with a massive particle in the center and 
a second particle placed randomly nearby. The refinement meshes were
constructed up to the specified level, so that both particles were located 
on this level. The usual potential and force calculations (described above)
were then performed to get the pairwise force between these two particles 
which was compared with the ``exact'' Newton force.\footnote
{We should note that particles 
in the CIC scheme (see \S 2) have a cubic rather than a spherical shape
so that the exact force at small separations is not an inverse square
function of distance.} Figure 6 shows the results of the force calculation 
with the FFT method (i.e., ART without refinements) and with
the relaxation method at different levels of refinement. We plot 
relative acceleration errors calculated as follows:
$$Error = \frac{\left\vert{\bf a}_{calc}\right\vert
-\left\vert{\bf a}_{theor}\right\vert}{\left\vert{\bf a}_{theor}\right\vert},$$
where $\left\vert{\bf a}_{calc}\right\vert$ is the acceleration calculated
on the mesh and 
$\left\vert{\bf a}_{theor}\right\vert$ is the theoretical acceleration. 
The upper panel in Figure 6 shows the relative force 
error versus interparticle separation, 
given in the units of the base grid, for a pure PM calculation and
for the second and fourth refinement levels. Note that despite the general 
similarity of the shape of FFT and relaxation forces, the scatter of the 
latter at small separations is larger than that of the former\footnote{
However, the magnitude of scatter 
is typical for PM codes. See, for example, Fig.1 in Villumsen (1989) or Fig.2b in Gelb (1992)}.
The scatter at small separations does not, however, 
mean that we have the same errors in particle orbits. 
We have performed a test placing two particles at the distance of 2 grid cells 
(where the scatter is largest) and giving them velocities in such 
a way that in 
case of a zero force error they would stay on a circular orbit. 
The deviation of 
the diameter of the orbit from its true value is a measure of 
the code accuracy. 
We performed the test for the PM force (two particles on the zeroth 
level at the 
distance of two base grid cells from each other) 
and relaxation force (two particles 
on the second refinement level 
at the distance of 2 second-level cells or 0.5 zeroth-level cells).
The resulting particle trajectories are shown in Figure 7. 
After more than two orbital periods the particles stayed 
on the nearly circular orbit with {\it maximum} error in diameter of about 8\%.
Note that we did not observe either significant drift of the orbit center or 
any catastrophic break of the trajectory. We conclude that even for separation 
equal to the resolution the code produces reasonably accurate trajectories of 
particles. 

The lower panel in Figure 6 shows the relative error for ART force 
calculated on the fourth level versus distance in units of 
the fourth-level mesh cell, together with the relative error 
corresponding to the Plummer softened force ({\em solid line}) with the 
softening parameter $\epsilon$ [$\phi\propto 1/\sqrt{r^2+\epsilon^2}$]
equal to the size of the fourth-level cell. 
Comparison shows that while the ART force error fluctuates around zero
for distances greater than two mesh cells,  the Plummer softened 
force is considerably weaker than $1/r^2$ law for up to six grid cells. 
The relation between resolution of the ART code (2 mesh cells of the 
maximum refinement level), $h_{ART}$,
and Plummer softening
length is thus $\epsilon\approx 3h_{ART}.$
Gelb (1992)
studied the shape of the Plummer softened force in his P$^3$M code. 
The result is consistent with ours (Gelb 1992, Fig. 2.4) -- 
the force starts to fall down at a distance about five times 
larger than the softening length. 
 
\subsection{Zeldovich pancake collapse}

One-dimensional plane wave collapse in an expanding universe
is one of the traditional tests of $N$-body codes 
(Klypin \& Shandarin 1983; Efstathiou et al. 1985). 
In this test, the analytic solution 
(Zeldovich 1970) is used to check how accurately the code 
integrates particle trajectories. If we know the initial 
conditions, the solution predicts particle positions for any other 
moment of time:
$${\bf x}_i^Z={\bf q}_i+\left(\frac{{\bf A}}{2\pi k}\right)
\cos \left(2\pi{\bf k}\cdot{\bf q}\right),$$
here ${\bf q}_i$ are the initial (unperturbed) positions, 
${\bf k}$ is the wavevector, and $A(t)$ is the amplitude. 
In an $\Omega =1$ universe, $A=a(t)/a_0$, where $a_0$ is a scale factor 
at the crossing time. The corresponding velocities can be obtained by 
differentiating the above formula for positions. 

We used a base grid of $32^3$ cells with $64^3$ particles. The particle
positions and velocities were initially perturbed using the Zeldovich 
approximation. The particle trajectories were integrated by the ART code
with three levels of refinement
(the density threshold for opening a new level was twenty particles
in a cell on the base grid and three particles in a cell for any refinement 
level). In Figure 8 we show one-dimensional phase diagrams at the 
crossing time that compare results of the ART code with results
of the PM code (grid of $32^3$ cells). 
The figure shows that the ART code follows 
the analytical solution more accurately than the PM code. This result is  
shown quantitatively in Figure 9, where we plot rms deviations of the particle
positions and velocities from the exact solution as a function of time
(Efstathiou et al. 1985):
$\Delta x_{rms}=\left[\sum\left(x_i-x_i^Z\right)^2/
\sum\left(x_i^Z-q_i\right)^2\right]^{1/2},$
$\Delta v_{rms}=\left[\sum\left(v_i-v_i^Z\right)^2/
\sum \left(v_i^Z)^2\right) \right]^{1/2}.$
 Starting from 
the moment when the first refinement level was created ($a\approx 0.29$)
the rms deviations were systematically lower in the ART code than in 
the PM code.

%---------------------------------
\subsection{Spherical infall test}
%---------------------------------

An analytic solution describing spherical infall of material onto an 
overdensity in an expanding cosmological framework was developed 
by Fillmore \& Goldreich (1984) and Bertschinger (1985). 
As noted by Splinter (1996), the problem possesses a symmetry different 
from the intrinsic planar symmetry of the mesh codes,
which makes it a useful and strong test. 
The analytic solution describes the evolution
of a spherical uniform overdensity $\delta\rho/\rho=\delta_i\ll 1$ 
in a region which has {\em proper} radius $R_i$ at some initial time $t_i$ in 
a flat Einstein-de Sitter universe. As the density contrast grows, 
the matter initially inside $R_i$ is increasingly decelerated. 
Eventually it stops expanding and turns around to collapse. 
If the initial Hubble flow is unperturbed (all peculiar velocities are 
initially equal to zero), the initial turnaround occurs at the 
time $t_{ita}$: 
$$t_{ita}=\frac{3\pi}{4}\delta_i^{-3/2}t_i.$$
At this moment the initial overdensity reaches its maximum radius:
$$r_{ita}=R_i\delta^{-1}_i.$$
Later on, shells of  successively larger and larger radii 
turn around. At a given time $t\gg t_{ita}$, a shell of a radius 
$$r_{ta}(t)=r_{ita}\left(\frac{t}{t_{ita}}\right)^{8/9}$$ 
starts to collapse (Bertschinger 1985).
 
The solution for the density profile of the overdensity is self-similar 
in terms of the dimensionless variable $\lambda\equiv r/r_{ta}(t)$ and 
for $\lambda\ll 1$ can be expressed as (Fillmore \& Goldreich 1984;
Bertschinger 1985):
$$D(\lambda)\approx 2.79\lambda^{-9/4},$$
while at larger $\lambda$ the main feature of the solution is the presence 
of sharp caustics. 

We modeled the spherical infall problem described above by distributing
$32^3$ particles uniformly on the $32^3$-cell grid (in the centers of
the grid cells) and placing additional particles in a sphere 
of radius 2 grid cell lengths in the center of the computational volume. 
The number of additional particles determines the initial overdensity,
which we have chosen to be $\delta_i =0.2$. 
We integrated particle trajectories from 
$a_i=t_i^{2/3}=0.1$ up to $a=10.1$, taking 10,000 steps to ensure that 
particles move only a fraction of a mesh cell
in a single time step on any of the five refinement 
levels, which is the condition 
required for the integration to be stable. The refinement levels were 
introduced in the regions where density was equivalent to more than six 
particles in a cell and the number of levels was limited to five,
thus making the effective resolution $1024$ in the highest density
regions. In Figure 10 the calculated density profile
is compared with the analytic solution (from Tables 4 and 5 of
Bertschinger 1985). We see a good agreement 
between the calculated density profile and the analytic 
solution at all radii down to the resolution limit. 

%----------------------------------------------------------------
\subsection{Realistic cosmological runs: comparison with PM code}
%----------------------------------------------------------------

To test the performance of the code in realistic cosmological simulations,
we made a set of runs using ART and PM codes with the 
same initial conditions and different spatial resolutions and compared 
the resulting particle distributions. 
 In the first four runs we simulated an $L=20h^{-1}$ Mpc box with 
$N=32^3$ particles, assuming a flat $\Lambda$CDM cosmological model 
($\Omega_{\Lambda}=0.7$, $h=0.7$, $\sigma_8=1.0$). In the ART runs we 
allowed for two levels of refinement starting from $64^3$ and 
$128^3$ grids -- we will call these runs ART $64^3+2$L and
ART $128^3+2$L. The number of time steps was $\sim 1000$ in 
the ART $64^3+2$L and $\sim 2000$ in the ART $128^3+2$L run. 
The refinement levels were introduced wherever the density exceeded 
a threshold value equivalent to more than 5 particles per cell. 
The PM code was run with $64^3$-cell (PM $64^3$) and $256^3$-cell 
(PM $256^3$) grids. 
In Figure 11 we show the projected particle distribution at $z=0$ from 
these four runs, with a subsets of particles belonging to one of the halos 
shown in a smaller window. 
The global distribution of particles and halos is well
reproduced by the ART code. Also, halos in the ART $64^3+2$L simulation
are much more compact than in the PM $64^3$ simulation. This result is shown 
quantitatively in Figure 12, where we compare density distribution functions
(the fraction of the total mass in the regions of a given 
overdensity) in these simulations.
The density distributions for all runs were computed after
rebinning the density field to the $256^3$ grid. The resolution of a 
simulation puts limits on the maximum density in the halo cores because 
gravitational collapse virtually stops at scales
of $\sim 1$ grid cell (e.g., Klypin 1996). 
Therefore, the density in the halo cores (the high-density tail of the 
distribution) is a good indicator of the spatial resolution. We note
that the density distribution functions for both the 
ART $64^3+2$L and the PM $256^3$ 
runs show approximately the same behavior, reaching overdensities of
$\approx 2\times 10^4$, while the PM $64^3$ run fails to produce halos with 
overdensities greater than $\approx 5\times 10^3$. We therefore conclude 
that the ART code produces density fields similar to those of a PM code 
of comparable resolution. 

The first application of the code was the study of the structure 
of dark matter halos. Therefore, as a final test we compared
the halo density profiles in the ART and PM simulations. 
The size of the simulation box, $L=15h^{-1}$ Mpc, was chosen to be the 
same as in the larger simulations 
described in the next section. The rest of the parameters were the same 
as in the above simulations. We simulated the evolution of the $32^3$ 
particles using the PM code with $256^3$-cell mesh and the 
ART code with a $64^3$-cell base
grid and three levels of refinement. As before, we refined 
regions where the local density exceeded a threshold value 
of about 5 particles
per cell. A halo-finding algorithm (described in the next section) was applied
to the resulting particle distribution. In Figure 13 we present the density 
profiles for six halos of different masses. The mass resolution in these
simulations ($1.2\times 10^{10} M_{\odot}$) determined the mass range 
of halos. The most massive halo in Figure 13 consists of about 5000 particles,
while the least massive contains only about 100 particles. The density 
profiles of PM and ART halos agree reasonably well down to the resolution 
limit ($\sim 60h^{-1}$ kpc).

%-----------------------------------------------------------------
\section{An application: structure of dark matter halos in CDM and 
         $\Lambda$CDM models}
%-----------------------------------------------------------------
\subsection{Motivation}
We used the code to study the structure of dark matter halos in two of 
the currently popular cosmological models:
standard cold dark matter (SCDM) and cold dark matter
with cosmological constant ($\Lambda$CDM) models.
The dark matter halos play a crucial role in the formation and dynamics 
of galaxies and galaxy clusters. Therefore, theoretical predictions
about the structural and dynamic properties of the halos can be compared 
with observations and used as a powerful test of a given theoretical model.
The numerical study of the halo structure requires very high spatial 
dynamic range (at least $\sim 10^4$) because the simulation box has to be 
large enough to account correctly for large perturbation waves 
and the force resolution has to be high enough 
to make predictions in the observational range ($\leq 5$ kpc). 
The ART code was designed to handle such high dynamic ranges.
 
The properties of dark matter halos were intensively investigated 
recently for a variety of cosmological models. Early numerical studies
(Frenk et al. 1985, 1988; Quinn, Salmon, \& Zurek 1986; Efstathiou et
al. 1988) indicated that the density profiles of dark 
matter halos in hierarchical clustering models in a flat, $\Omega=1$, universe
were approximately isothermal [$\rho(r)\propto r^{-2}$], in agreement with 
analytic results (Fillmore \& Goldreich 1984; Bertschinger 1985). The 
dependence of the halo density profiles on the initial perturbation spectrum 
and on specific parameters of the cosmological model were also 
studied both analytically (Hoffman \& Shaham 1985; Hoffman 1988) 
and numerically (e.g., Crone, Evrard, \& Richstone 1994). 
These early numerical studies, however,  
lacked the necessary mass and spatial resolution to make reliable 
predictions on the structure of the halo cores. 
To overcome the resolution limits, substantial 
efforts were made to simulate the formation of halos from 
isolated density perturbations (e.g., Dubinski \& Carlberg 1991; Katz 1991) 
or to resimulate with a higher resolution halos identified in large 
low-resolution runs (Navarro, Frenk, \& White 1996a, hereafter NFW; 
Tormen, Bouchet, \& White 1997). 
These simulations have the advantage of simulating halos 
in a wide mass range with  homogeneous spatial and mass resolutions. 
However, it is hard to infer the statistically reliable results from these 
simulations because only a few halos are simulated. 
In a direct simulation one can get
a statistically significant sample of halos suited for more detailed analysis.
Studies of the structure and dynamics of halos extracted from high-resolution 
direct simulations were done by Warren et al. (1992) and 
Cole \& Lacey (1996). Warren et al. (1992) used a TREE code to simulate 
the evolution of $128^3$ particles in an $\Omega=1$ universe 
with scale-free initial conditions. The force resolution was determined 
by imposing a Plummer softening of $\epsilon=5$ kpc. 
Special emphasis was given to the investigation of halo 
shapes. Similar initial conditions were used in the study 
by Cole \& Lacey (1996), who used a P$^3$M code to evolve $128^3$ particles. 
The resolution in the latter simulations was $L/\epsilon=3840$, 
where $L$ is the size of the computational volume
and $\epsilon$ is the Plummer softening parameter. The results 
indicated that the density profiles of {\em all} simulated halos are well 
fit by the analytical model of NFW. 
This model has $\rho\propto r^{-1}$ at small radii and steepens smoothly to 
$\rho\propto r^{-3}$ at a {\em scale radius} $r_s$:
\begin{equation}
\rho(r)\propto \frac{1}{r(1+r/r_s)^2}.
\end{equation}
The density profile described by this expression is singular,
because the density rises arbitrarily high when $r \to 0$, forming a cusp. 
The cuspy structure of the central parts of a halo thus  represents  
a generic prediction of the model. 
Although the NFW profile is consistent with current X-ray and 
gravitational lensing observations of galaxy clusters (NFW), 
the $\rho \propto r^{-1}$ behavior is in contradiction with observations of 
dynamics of dwarf spiral galaxies that imply flat central density profiles 
(Flores \& Primack 1994; Moore 1994; Burkert 1996). 
These observations can serve as one of the critical tests of any model that 
includes a dark matter component because it is generally believed that 
the dynamics of the dwarf spiral galaxies is dominated by dark matter 
on scales $r\geq 1$ kpc. The fact that the NFW profile holds for 
a variety of cosmological models (NFW; Cole \& Lacey 1996) indicates its 
possible universality for CDM-like models. The goal of the present study 
was to investigate the structure of dark matter halos  formed in 
a $\Lambda$CDM model. This model is currently one of the most successful
scenarios of structure formation in the Universe. It is, therefore,
important to check whether the central cusp is present in halos formed 
in this model. 
%-----------------------
\subsection{Simulations}
%-----------------------

To study the structure of dark matter halos, we simulated 
the evolution of $64^3$ particles in standard CDM ($\Omega=1$, $h=0.5$, 
$\sigma_8=0.63$) and $\Lambda$CDM ($\Omega=0.3$, $\Omega_{\Lambda}=0.7$,
$h=0.7$, $\sigma_8=1.0$) models. We made three runs: one high-resolution
(resolution $\sim 2h^{-1}$ kpc) run for each models, and 
a lower resolution run (resolution $\sim 8h^{-1}$ kpc) for the $\Lambda$CDM 
model to study the effects of resolution. 
In terms of the Plummer softening length (see \S 4.1), 
our resolution corresponds to $L/\epsilon \approx 12000$ for the high 
resolution runs, and $L/\epsilon \approx 3000$ for the low-resolution
run. The simulations were started at $z=30$ and 
the particles trajectories were integrated by taking 3872 time steps 
in the low-resolution run, and 7743 time steps in the high-resolution runs. 
The size of the simulation box, $L=15h^{-1}$ Mpc, 
determined the mass resolution (particle mass) as
$3.55\times 10^9h^{-1} M_{\odot}$ for SCDM and $1.06\times 10^9h^{-1} 
M_{\odot}$ for $\Lambda$CDM. 

%----------------------------------
\subsection{Halo-finding algorithm}
%----------------------------------
To identify halos in our simulations, we use an algorithm similar to that 
described in Klypin, Primack, \& Holtzman (1996). The algorithm 
identifies halos as local maxima of mass inside a given radius. 
The efficiency of the algorithm was improved by incorporating the 
idea of Warren et al. (1992)
of finding approximate locations of density peaks using particle accelerations.
This idea is based on the principle that particles 
with the largest accelerations 
should reside near the halo centers, which is true for halos with roughly
isothermal density profiles. 
In practice, this way of finding density maxima has proved to be quite 
efficient. The halo identification algorithm can be described 
as follows.

1. The particles are sorted according to the magnitude of their scalar 
   accelerations. 

2. The particle with the largest acceleration determines the approximate
   location of the first halo center. Particles located inside a sphere
   of radius $r_{init}$ centered at the halo center 
   are assigned to the same halo and are excluded
   from  the list of particles used to identify halos. The radius $r_{init}$ 
   is an adjustable parameter; we use a radius approximately twice as large
   as the force resolution of a simulation. The procedure repeats for 
   the particle 
   with the largest acceleration in the list of remaining particles. 
   The peaks are identified until there are no particles in the list.

3. When all the density peaks are identified, we proceed to find more 
   accurate positions of the halo centers. This is done iteratively by 
   finding the center of mass of all particles inside $r_{init}$ and 
   displacing the center of the sphere to the center of mass. 
   The procedure is iterated until convergence. 

4. When the halo centers are found, we increase $r_{init}$ 
   until the overdensity inside the corresponding sphere reaches a certain 
   limit. The limit is based on the top-hat model of gravitational collapse, 
   which predicts the typical overdensity for virialized objects $\sim 200$ in
   CDM and $\sim 334$ for our $\Lambda$CDM model (e.g. Lahav et al. 1991; 
   Kitayama \& Suto 1996). However, we denote halo radius and the 
   mass inside this
   radius defined as $M_{200}$ and $r_{200}$ regardless of the actual value
   of the limit. Smaller halos located within a radius 
   $r_{200}$ of a bigger halo are deleted from the list.

As an output we get a list of halo positions, velocities, and parameters 
(such as $r_{200}$ and $M_{200}$). 

%------------------------------------------
\subsection{Results: Halo Density Profiles}
%------------------------------------------
We applied the halo-finding algorithm described above to identify 
halos in the simulations at zero redshift. 
Only halos with more than 100 particles within $r_{200}$ 
were taken from the full list. We also present results 
for relatively isolated halos, excluding all halos 
which have close ($r<2r_{200}$) neighbors of mass more than
half of the halo mass.  
 The density profiles were  constructed by estimating the density 
in concentric spherical
shells of logarithmically increasing thickness, with the smallest radius 
corresponding to the maximum resolution. 
The resulting profiles were fitted with the analytical formula of NFW 
(eq. [19]), taking the the scale radius $r_s$ as the fit parameter. 
In Figures 14 and 15 we show density profiles 
of nine halos of different mass identified in the high-resolution CDM and 
$\Lambda$CDM simulations, along with the analytic fits to the halos. 
The NFW profile appears to be 
a good approximation for halos of all masses (within the mass range of 
our simulations) in both CDM and $\Lambda$CDM
models. 

NFW argued that the {\em concentration parameter}, $c=r_{200}/r_s$ 
(see eq.[19]), of a dark matter halo depends on the halo mass. 
They found that low-mass halos are more centrally 
concentrated than high-mass ones, which possibly reflects 
a trend in the formation redshifts of halos. Figure 16 shows 
the concentration $c$
as a function of halo mass ($M_{200}$) for halos identified in our 
simulations.
The solid curve represents a theoretical prediction (see NFW for discussion)
assuming the definition
for the formation time of a halo as the first time when half of its
final mass $M_{200}$ lies in progenitors with individual masses
exceeding a fraction $f=0.01$ of $M_{200}$. This particular value of $f$
seemed to provide the best approximation to the numerical results of NFW for 
the CDM model. 
The results of both the CDM and the $\Lambda$CDM simulations 
agree reasonably well with this curve. The larger spread of parameter
$c$ for low-mass  halos arises mainly from 
statistical noise. The lowest mass halos 
($M_{200}\sim10^{11.5}$ M$_{\odot}$) contain a few hundred particles 
within their $r_{200}$ and thus have more noisy 
density profiles (typical $2\sigma$ error in $\log c$ $\sim 0.2-0.3$)
than more massive halos 
($M_{200}>10^{13}$ M$_{\odot}$) which have tens of thousands of particles
(error in $\log c$ $\sim 0.05$). 
%---------------------------------
\subsection{Effects of resolution}
%---------------------------------
It is important to model reliably the structure of the very central part
($r<10h^{-1}$ kpc) of a halo because that is the part 
for which we can compare model 
predictions with observational results. Unfortunately, that part is also where
force resolution may strongly affect the shape of the density profiles. 
To study possible effects of force resolution, 
we have compared density profiles 
of {\em the same} halos taken from $\Lambda$CDM simulations of different 
resolution described in \S 5.2. In Figure 17 we compare density profiles
of four halos from these simulations. As before, the density profile 
is drawn only to the resolution limit of the simulation. We conclude that,
up to the resolution limit, the lower resolution density profile follows
the higher resolution density profile. 

%-----------------------
\subsection{Conclusions}
%-----------------------
We studied the structure of dark matter halos in 
CDM and $\Lambda$CDM models with a resolution of $\sim 2h^{-1}$ kpc in 
a box of $15h^{-1}$ Mpc. 

1. We found that for $r<r_{200}$, the density profiles of {\em all} halos 
in both CDM and $\Lambda$CDM simulations
are well fitted by the analytical formula (eq. [19]) of NFW. 

2. The mass dependence of the halo  concentration parameter $c$
 in our simulations is consistent with the results of NFW. 

3. The fact that our results for the CDM model agree with the results 
of NFW serves both as a final test of the presented 
code and as an independent check of their method with results from 
{\em direct} cosmological simulations. 
%------------------------------------
\section {Discussion and conclusions}
%------------------------------------
We present a new high-resolution $N$-body code that incorporates 
the idea of an adaptive refinement tree (Khokhlov 1997) 
to build a hierarchy of refinement meshes in regions where higher resolution 
is desired. Unlike other $N$-body codes that make use of refinement
meshes, our code is able to construct meshes of arbitrary 
shape covering both elongated structures 
(such as filaments and walls) and roughly spherical dark 
matter halos equally well. 
The meshes are {\em modified} to adjust to the evolving 
particle distribution instead of being {\em rebuilt} at every time step. 
We use a cubic grid as the zeroth level of the mesh hierarchy. 
The size of this grid determines the minimum possible resolution of
a simulation (i.e., resolution in regions where there are no refinements).
The code blocks working on the zeroth-level grid are similar to those
of a PM code. 
To solve the Poisson equation on refinement meshes, we have 
developed a new solver that uses a multilevel relaxation method with 
successive overrelaxation (Hockney \& Eastwood 1981; Press et al. 1992). 
The solver is fully parallel and an FFT solver is only twice as fast as
our solver for the same number of mesh cells. In real simulations with 
the same resolution, the relaxation solver outperforms the FFT because 
the resolution is achieved with a much smaller number of cells 
(see \S 3.7).
The tests  presented (\S 4) show that our code 
adequately computes gravitational forces down to scales of 
$\sim 1.5-2$ mesh cells. 
The memory overhead in the current 
version of the code is rather large compared to that of other high-resolution 
codes. The number of required mesh cells, however, changes very slowly
with increasing resolution. At present, the code is capable 
of handling a dynamic range of $\sim 10,000$ and higher. 
In our latest runs, for which we have used modified 
version of the code incorporating multiple time steps,
 we reached a dynamic range of
$\sim 24,000$ for a system of $128^3$ particles.  Tests of the code 
performance show that it is about three times faster than an AP$^3$M code 
(and, therefore, TREE code; see Couchman 1991) of comparable resolution. 
Still, the condition requiring that particles move only a certain 
fraction of a mesh cell at every time step makes the code CPU rather than 
memory limited. 

The present version of the code is by no means optimal 
and we plan the following improvements.

1. The memory requirement of the code can be significantly reduced 
   if pointers that are used to support the tree refinement structure are 
   shared by siblings (descendants of the same parent cell). The 
   memory overhead can be reduced even further by incorporating 
   more elaborate data-storage algorithms (Khokhlov 1997). The data 
   structures can also be changed to allow for parallelization on 
   distributed memory architectures.

2. The version of the code presented, like any other 
   high-resolution code, is constrained by the condition that 
   particles move only a fraction of a mesh cell
   in a single time step on any of the refinement levels. 
   To ensure that this condition is satisfied on the maximum 
   refinement level requires small time steps redundant 
   for particles moving on coarser meshes. We are now working on 
   integration scheme with multiple time steps. 
 
3. We plan to integrate the present $N$-body code with a high-resolution 
   Eulerian hydrodynamics code (Khokhlov 1997) that works on 
   similar refinement meshes. 

We have used the ART code to study the structure of dark matter 
halos in two cosmological models -- standard CDM ($\Omega=1$, $h=0.5$, 
$\sigma_8=0.63$) and a variant of $\Lambda$CDM 
($\Omega=0.3$, $\Omega_{\Lambda}=0.7$, $\sigma_8=1.0$). 
We have found that halos formed in $\Lambda$CDM model have 
density profiles similar to halos formed in CDM model.
The density profiles are well described by the analytical formula
(eq.[19]) presented by Navarro et al. (1996a) and have cuspy
[$\rho(r)\propto r^{-1}$] structure in the central ($r<10h^{-1}$ kpc)
parts of a halo with no indications of a core down to the resolution 
limit of our simulations.  
The similar model with different parameters
($\sigma_8=0.7$ and $h=0.5$), whose are not ideal because of rather small 
Hubble constant, was independently studied in recent papers by 
Navarro (1996) and Navarro, Frenk, \& White (1996b)
with conclusions different from ours. These authors conclude 
that halos formed in $\Lambda$CDM model are less centrally concentrated and
are thus more in accord with dynamics of dwarf galaxies. Our analysis 
has shown that the major source of this incosistency lies in the 
definitions of 
halo radius and concentration parameter $c$: in the above studies the authors 
neglect the fact that virialization radius in the $\Lambda$CDM model 
corresponds to overdensity $334$ rather than to $200$; they also normalize 
their densities to the critical density rather than average density of the 
universe (as assumed in the top hat collapse model). We were able to 
reproduce their results when we followed their definitions. 
 We therefore conclude that halos formed 
in the $\Lambda$CDM model have structure similar to that of the CDM halos
and thus cannot explain the dynamics of the central parts of dwarf 
spiral galaxies inferred from the galaxies' rotation curves.
\acknowledgements
We would like to thank Almadena Chtchelkanova (Berkeley Research 
Associates, Inc.) for useful advice 
on various data structures and algorithms. We would like also to thank 
Gustavo Yepes, Michael Norman, and Michael Gross for fruitful discussions
on various aspects of numerical simulations. We are grateful to 
Chris Loken for help in improving the presentation of the paper. 
This project was supported in part by the grant AST9319970 from the 
National Science Foundation and by the Naval Research Laboratory 
through the office of Naval Research. The simulations were done at the 
National Center for Supercomputing Applications (NCSA). 
%\vskip 10cm
%----------------------------------------------------------------------------

%
\newpage
\begin{figure}
\plotone{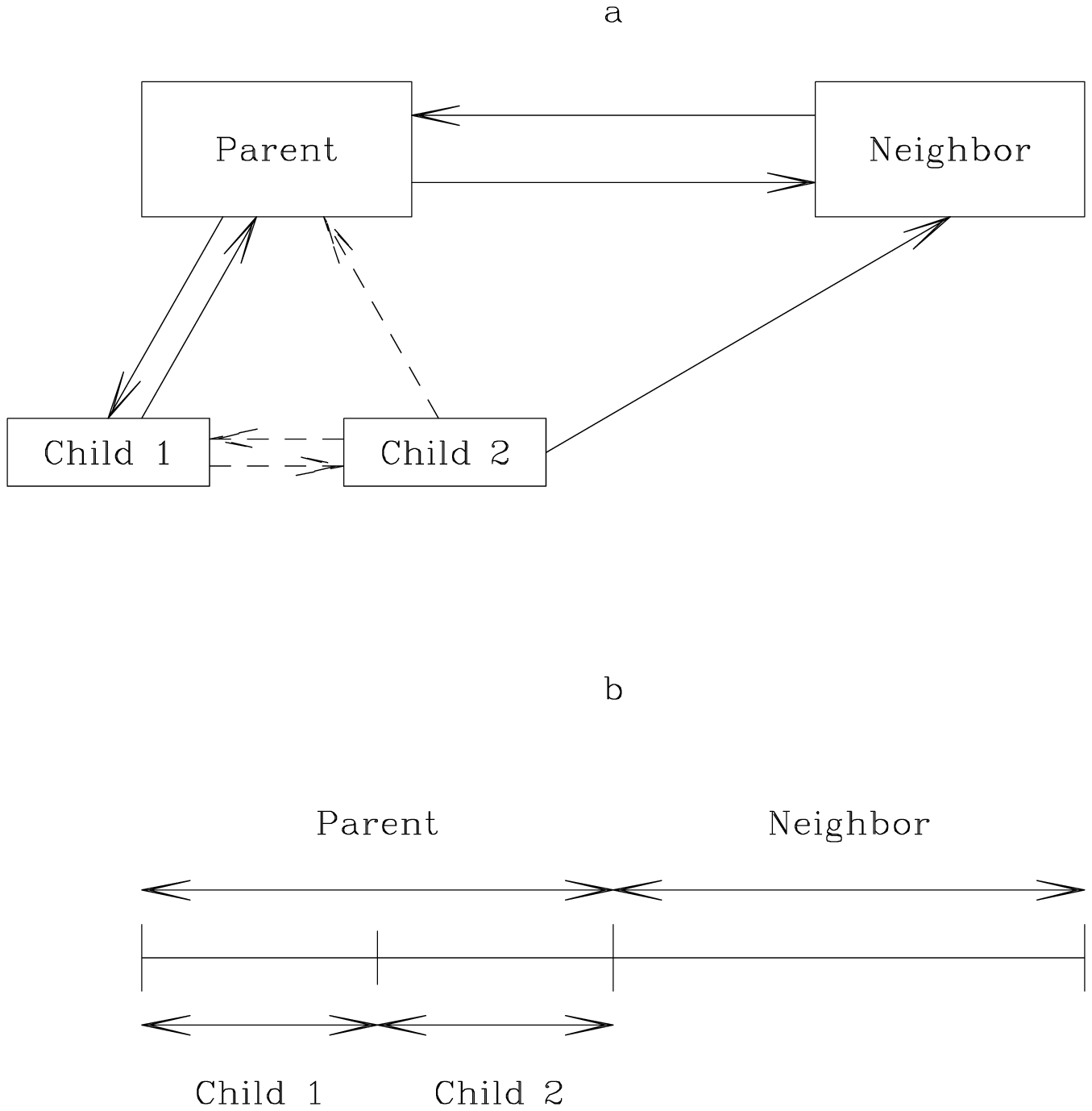}
\caption{ Schematic illustration of the tree structure and 
the pointers used to support it in one dimension. 
(a): Two neighbor cells (``Parent'' and ``Neighbor''), 
one of which (``Parent'') 
is split (it has two children marked ``Child1'' and ``Child 2'') 
and the other of which (``Neighbor'') is a leaf. The arrows denote pointers
used to support the structure (see text for details). Arrows 
drawn with dashed lines denote pointers which can be shared 
by all children. Below (b) we show the actual locations of the 
mesh cells in space.} \label{tree structure}
\end{figure}
\begin{figure}
\plotone{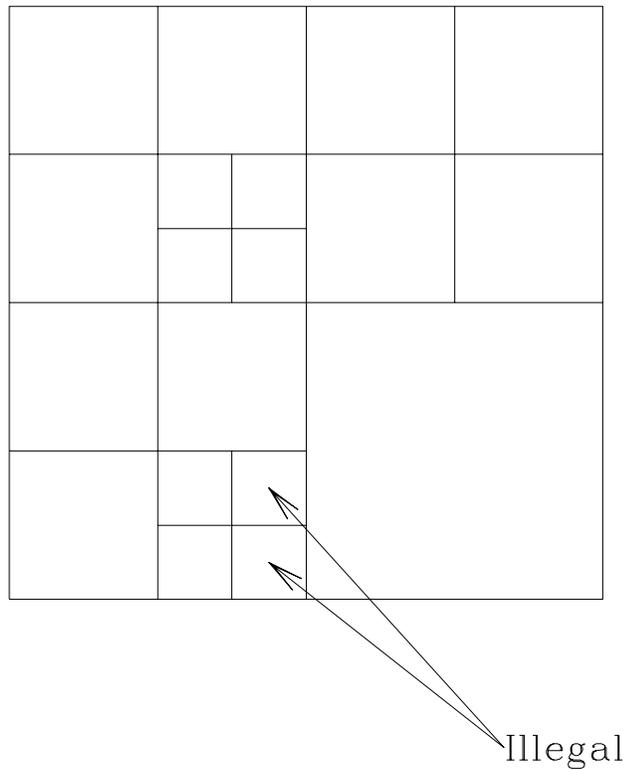}
\caption{Examples of permitted and prohibited neighbor 
configurations. The prohibited configurations (the cells that
violate the neighbor rule) are indicated with 
arrows. Note that a mesh is not allowed to have neighbors with level 
difference greater than 1 but it is allowed to have ``corner neighbor'' 
with level difference 2.} \label{illegal}
\end{figure}
\begin{figure}
\plotone{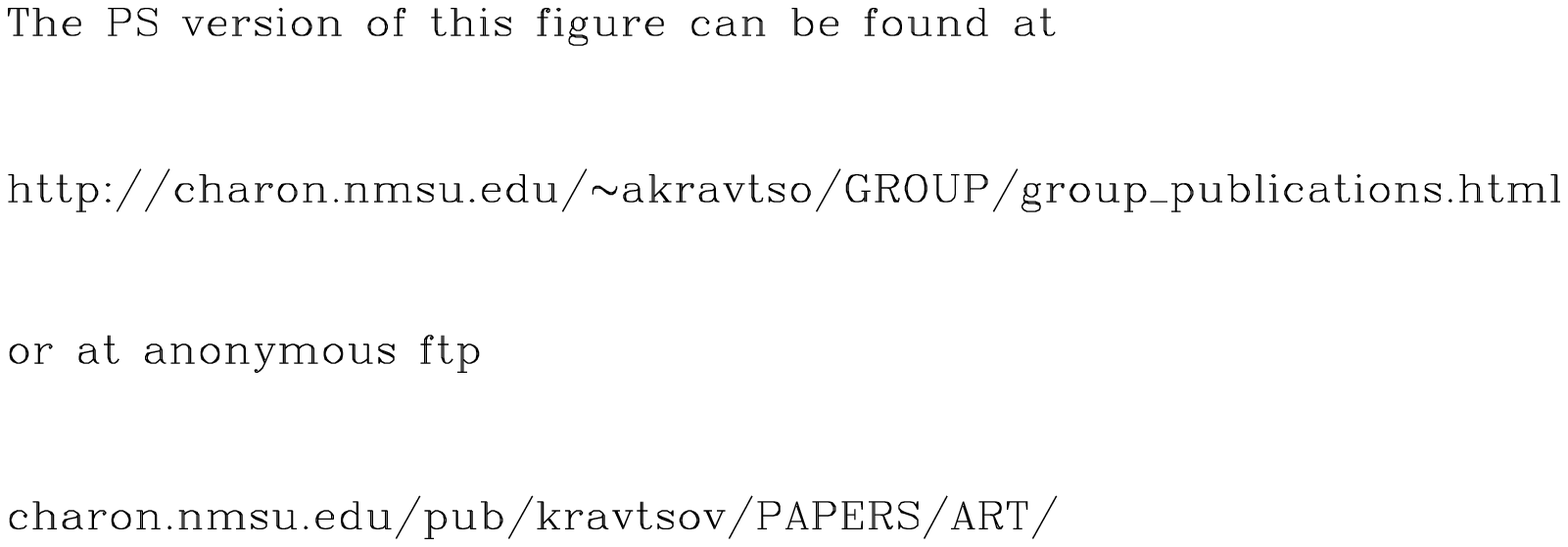}
\caption{(a) A slice through the refinement structure 
(base grid is not shown) in one of the $\Lambda$CDM simulations 
with $32^3$ particles 
(see \S 4.4)  and (b) the corresponding slice through 
the particle distribution. 
The area enclosed by the square in (a) is enlarged in Fig.4} \label{mgp}
\end{figure}
\begin{figure}
\plotone{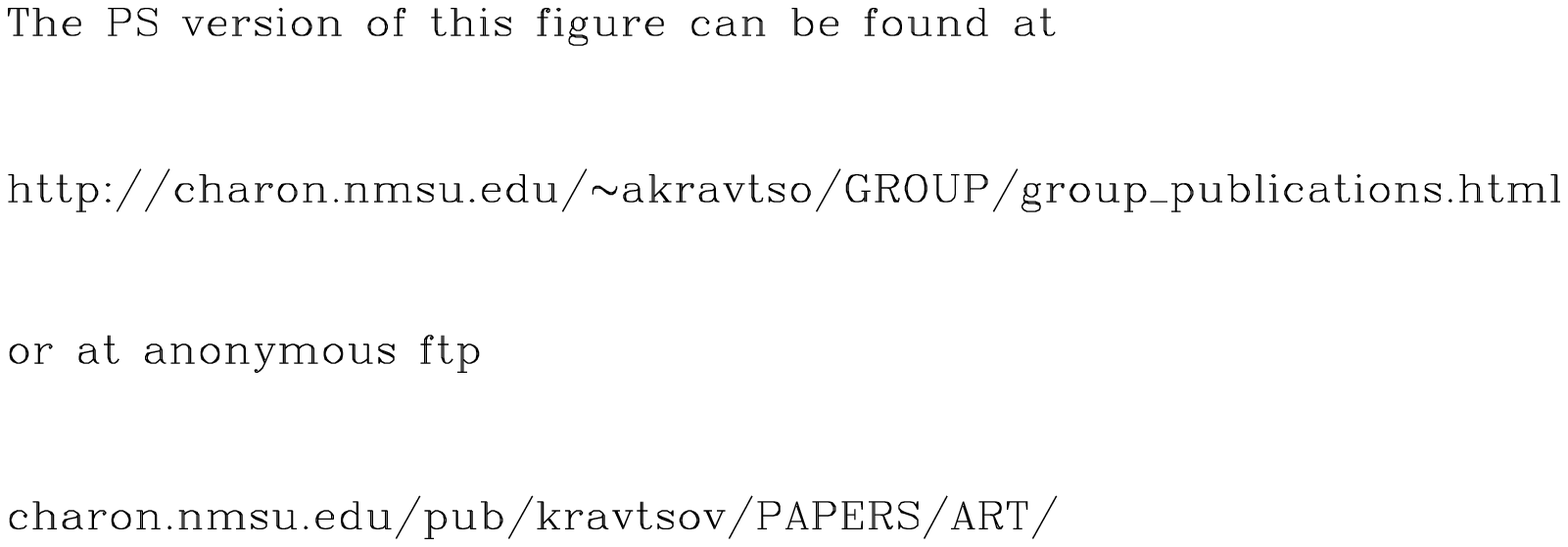}
\caption{Region enclosed by the square in Fig. 3a. 
Note that the mesh generator tends to build almost rectangular meshes around 
dense isolated clumps of particles, while to trace a filament
the generator creates meshes of arbitrary shape that effectively 
cover the elongated structures in particle distribution.
} \label{mgp_blow}
\end{figure}
\begin{figure}
\plotone{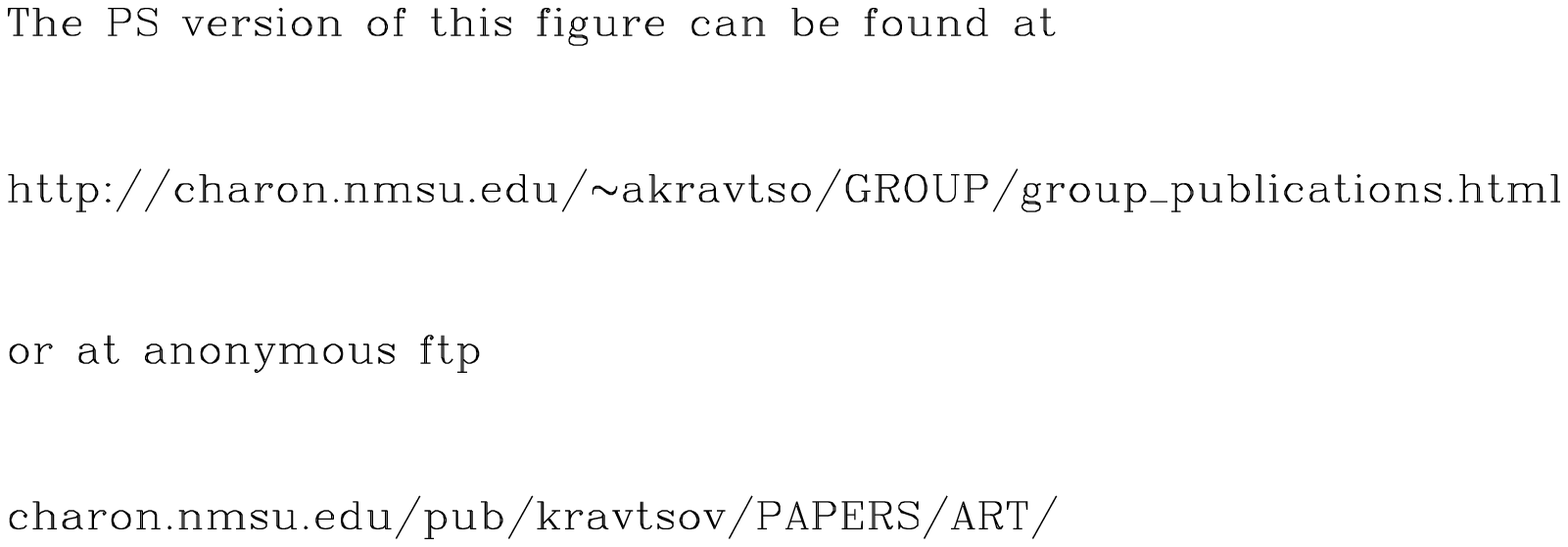}
\caption{ (a) Timing for $\Lambda$CDM run with $32^3$ particles
on 8 processors
of SPP-1200 Exemplar (see \S 3.8 for discussion). The total CPU time 
per step is scaled down by a factor of 2 for convenience. 
(b) Energy conservation error vs. expansion factor $a$ 
in $\Lambda$CDM runs with $32^3$ (solid line) and
$64^3$ (dashed line) particles.} \label{timing}
\end{figure}
\begin{figure}
\plotone{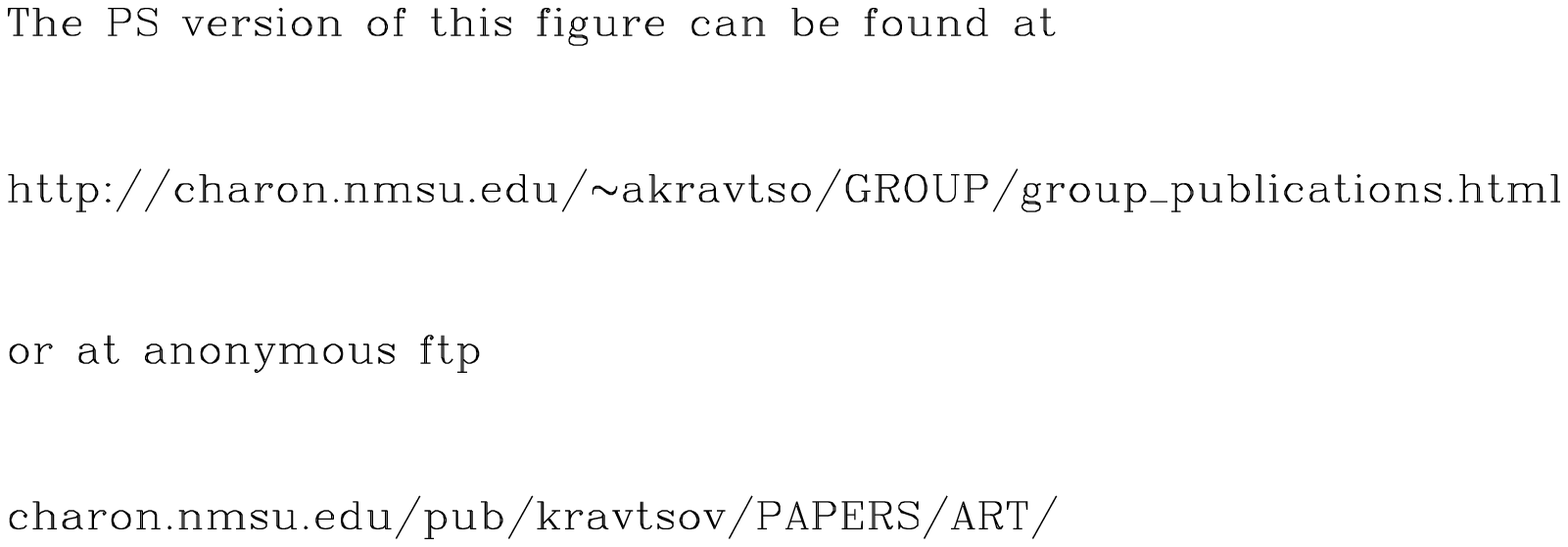}
\caption{(a) Pairwise force accuracy of the ART code on the base 
regular grid (FFT solver) and on the second and fourth refinement 
levels (relaxation solver) 
vs. interparticle separation. (b) Comparison of the force accuracy 
on the fourth refinement level with the theoretical accuracy of a Plummer 
softened force, both vs. interparticle separation in units of the
fourth-level cell length. 
} \label{force}
\end{figure}
\begin{figure}
\plotone{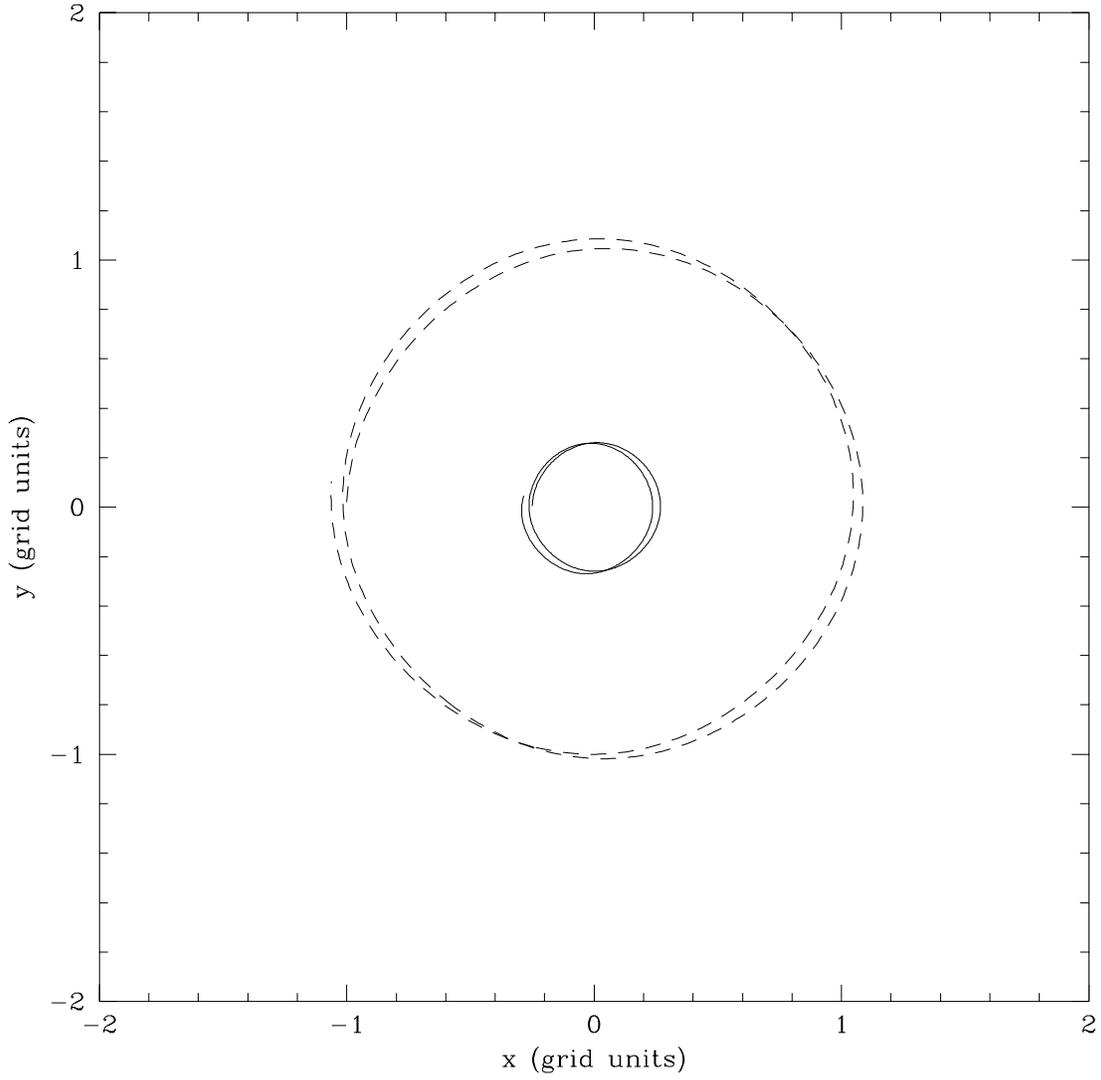}
\caption{ The circular motion test.
Two particles were placed 2 grid cell lengths apart and were given
velocities in such a way that in case of a zero force error they would stay on
a circular orbit. The dashed line shows trajectory (two orbital periods)
of two particles moving on the zero PM level (at the distance of two base grid 
cells from each other) and the solid line shows the trajectory of
particles moving 
on the second refinement level (at the distance of 2 second-level cells or 0.5 
zeroth-level cells).
After more than two orbital periods, the particles stayed 
on the nearly circular orbit with {\it maximum} error in diameter of about 8\%.
} \label{circtest}
\end{figure}
\begin{figure}
\plotone{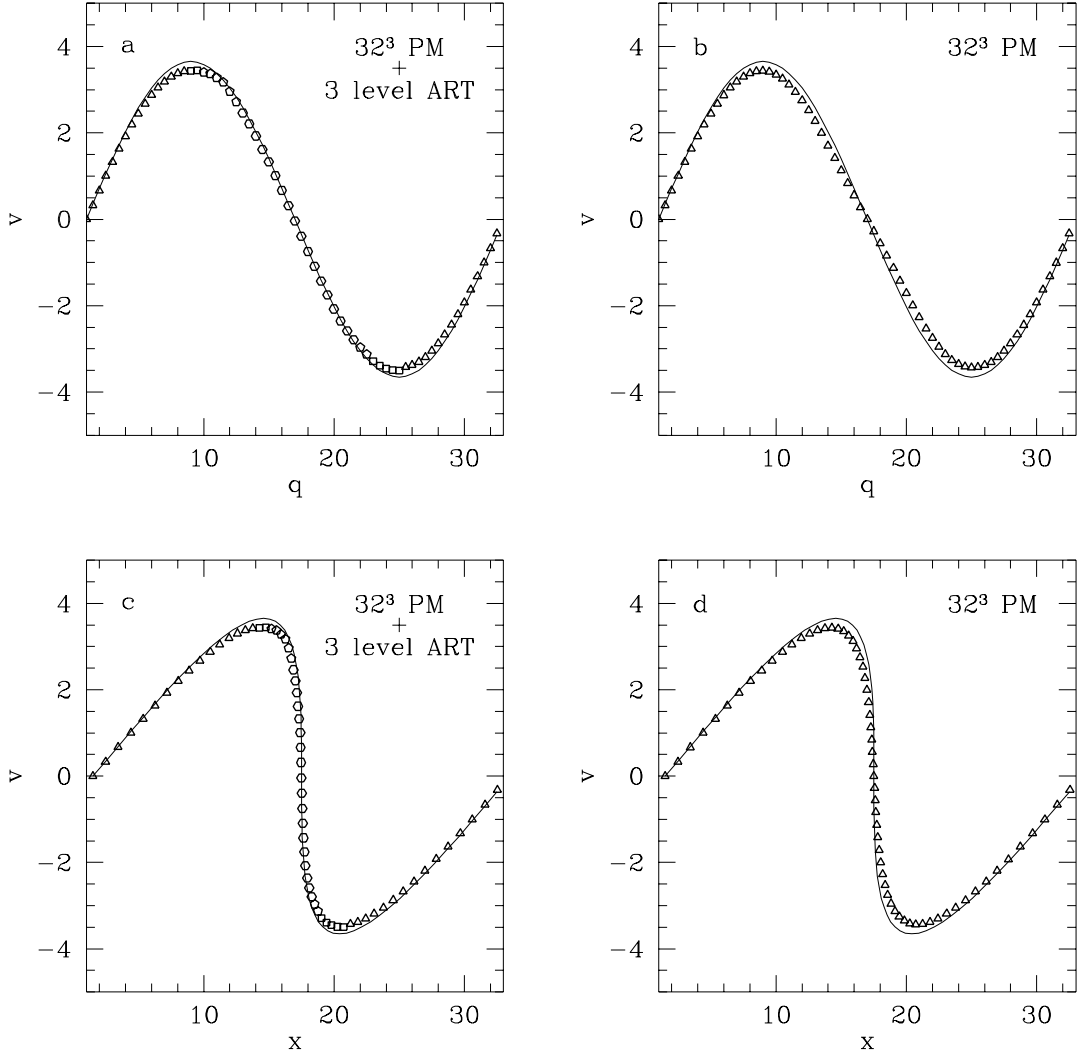}
\caption{Plane wave collapse test: 
phase diagram in Lagrangian coordinates $q$ (a)
 with 3 refinement levels and (b) in the PM  
simulation with a $32^3$-cell grid at the crossing time.
{\em Solid line}, analytic solution; {\em polygons}, numerical 
results. The number of polygon vertices is equal to the mesh level 
plus three, so that triangles show particles located on the base grid, 
squares show particles located on the first 
refinement level, and so on. (c,d) Corresponding phase diagram for physical 
coordinates $x$. 
The Lagrangian coordinates show the differences between 
ART and PM results more clearly, because at the crossing time 
the $v-x$ phase diagram is almost vertical in the central part of 
the pancake.}
\label{ptest_xv}
\end{figure}
\begin{figure}
\plotone{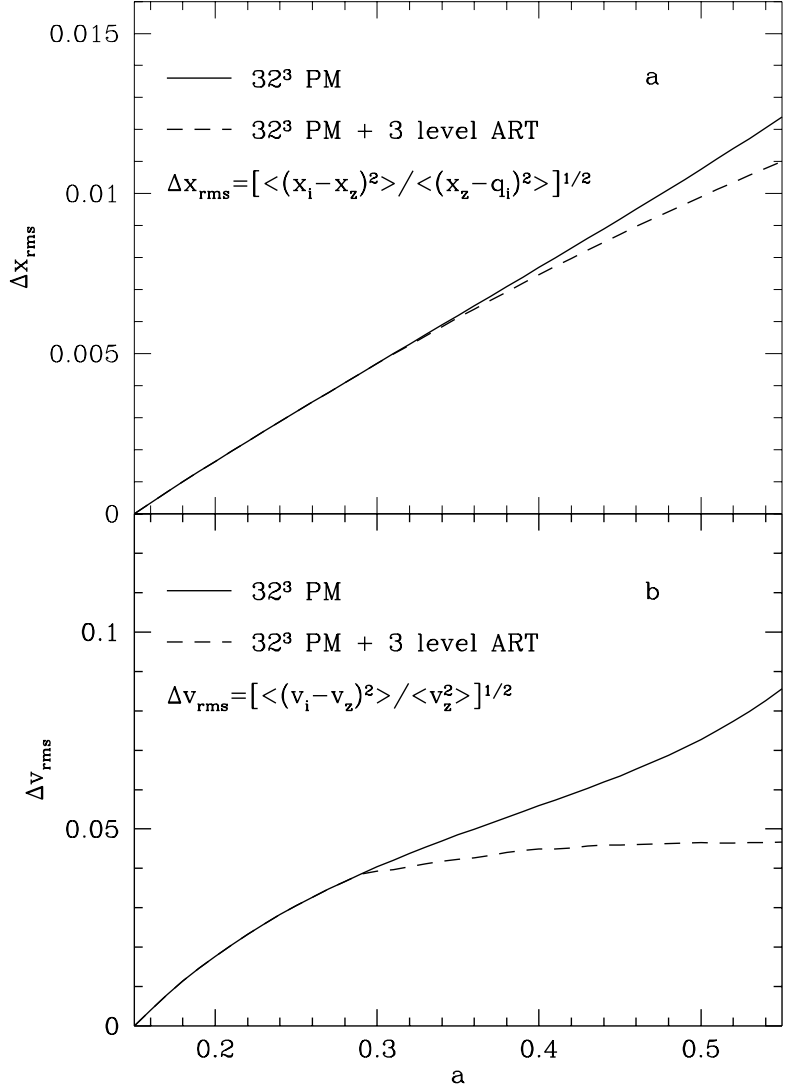}
\caption{Plane wave collapse test: rms deviations of (a) coordinates 
and (b) velocities from 
analytic solution vs. the expansion parameter for the PM code ({\em solid line})
and for the ART code with three levels of refinement ({\em dashed
  line}). Starting 
from the moment when the first refinement level was introduced 
($a\sim 0.29$),
the rms deviations of the ART code were lower than those of the PM code} 
\label{ptest_rms}
\end{figure}
\begin{figure}
\plotone{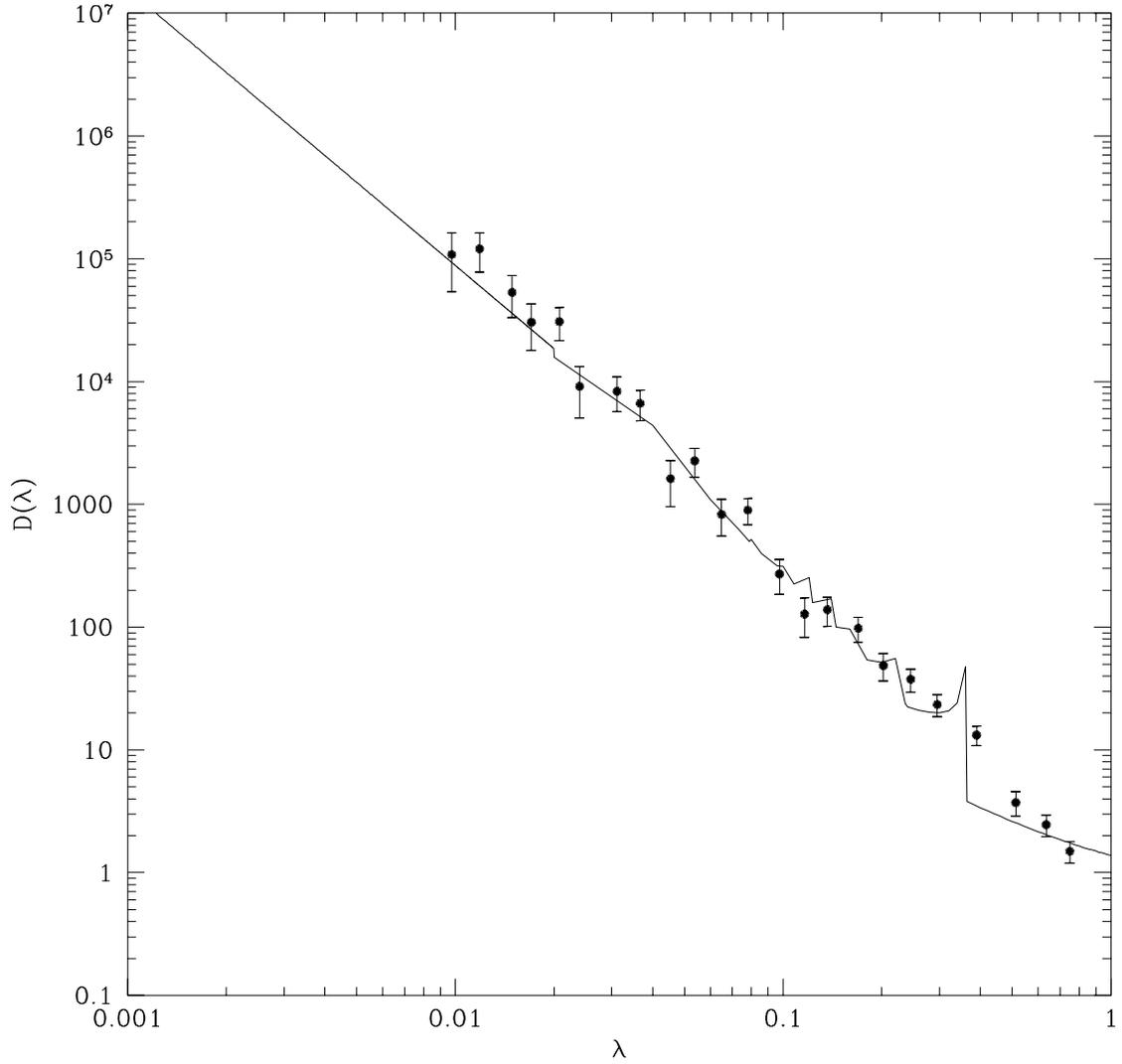}
\caption{Spherical infall test: the simulated density profile ({\em filled
circles}) compared with the analytic solution ({\em solid line}). 
The points for the analytic solution are 
taken from Tables 4 and 5 in Bertschinger (1985). The simulated density
profile was constructed by estimating the density in concentric spherical
shells of logarithmically increasing thickness, with the smallest radius 
corresponding to the maximum resolution. The error bars correspond to 
the Poisson noise. 
} 
\label{bert}
\end{figure}
\begin{figure}
\plotone{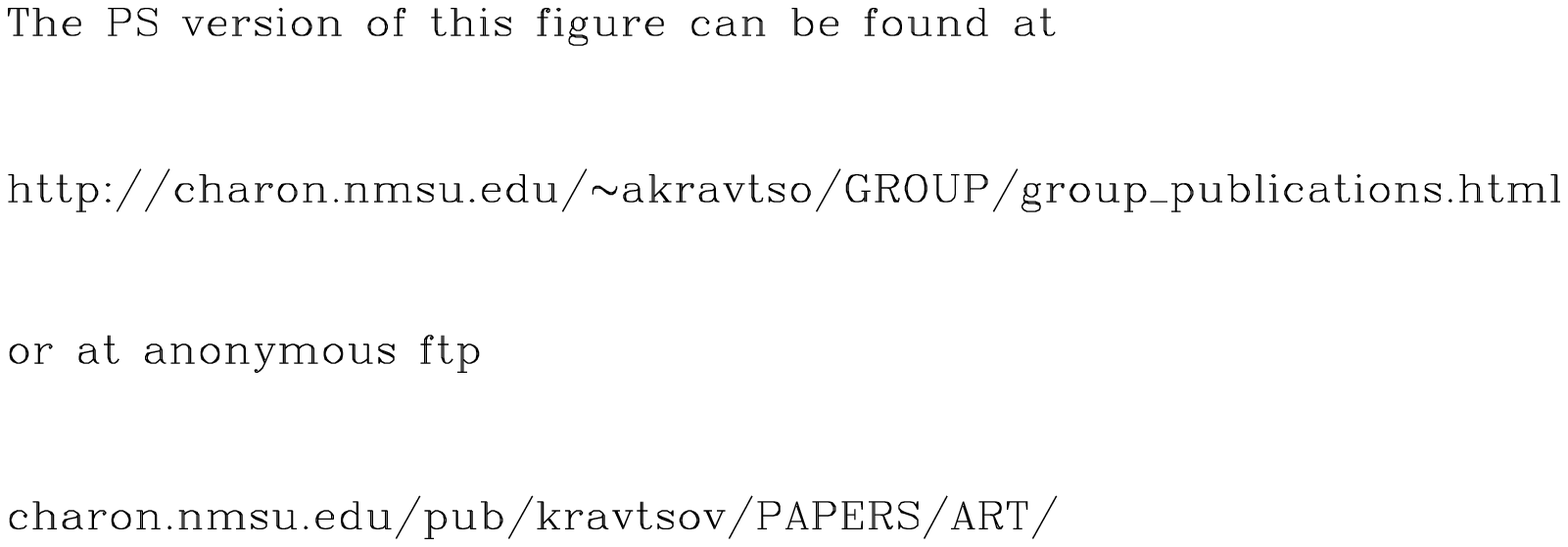}
\caption{Comparison of projected final distributions of $32^3$
particles in pure PM runs with (a) $64^3$-cell and (d) $256^3$-cell 
grids and in 
ART runs with two levels of refinement and base grid of (b) $64^3$
and (c) $128^3$ cells. In the small windows we show subsets of particles 
belonging to one of the halos (the same halo for all runs). 
} 
\label{prj4}
\end{figure}
\begin{figure}
\plotone{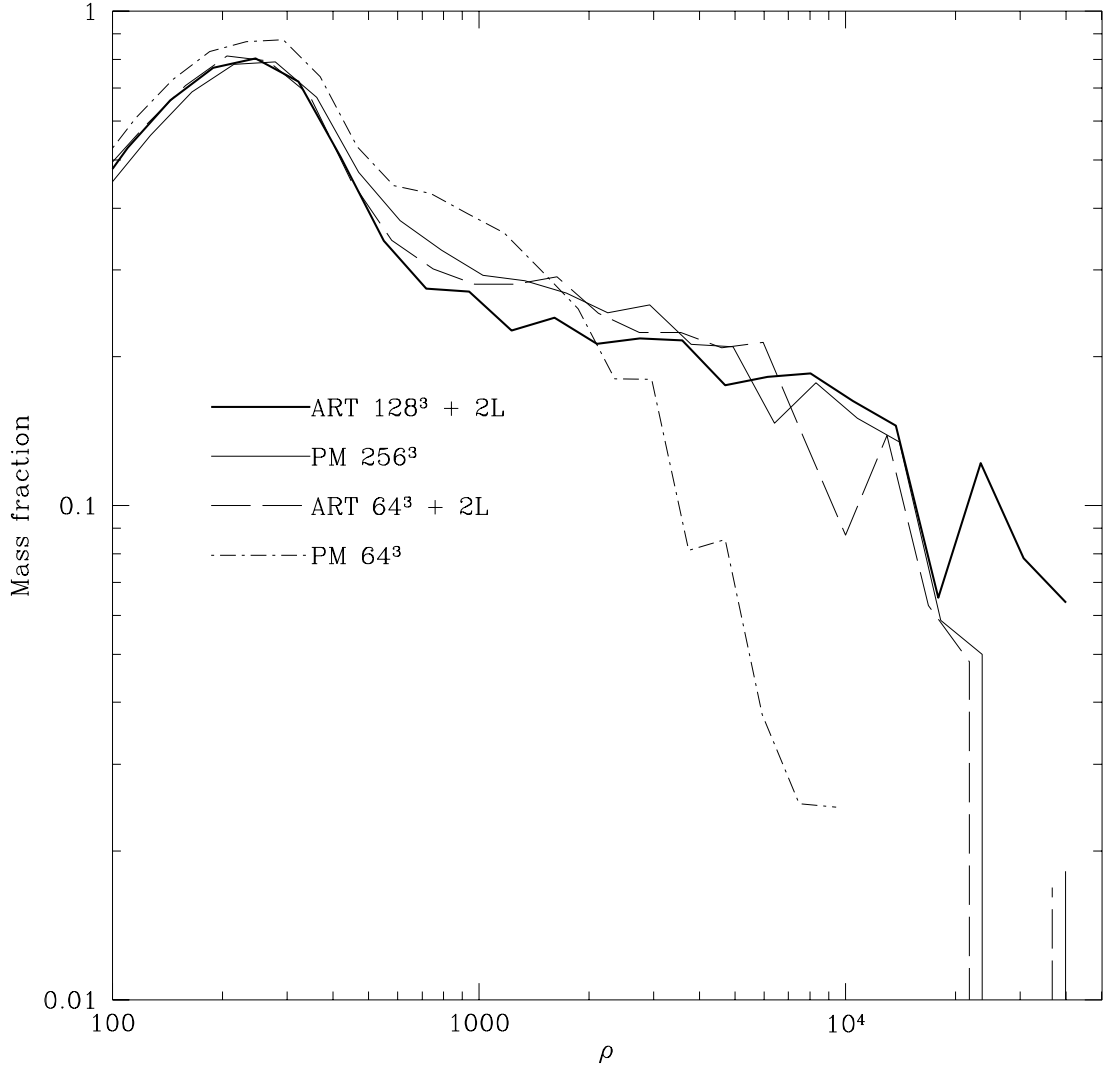}
\caption{Comparison of density distributions (fractions
of the total mass in the regions of a given overdensity) for 
PM and ART (two refinement levels) runs. 
The density distributions for all runs were computed after
rebinning the density field to the $256^3$-cell grid.  Note
that the density distribution functions for both ART $64^3+2$L and PM $256^3$ 
runs show approximately the same behavior, reaching overdensities of
$\approx 2\times 10^4$, whereas the PM $64^3$ run fails to produce halos with 
overdensities greater than $\approx 5\times 10^3$.
} 
\label{rho}
\end{figure}
\begin{figure}
\plotone{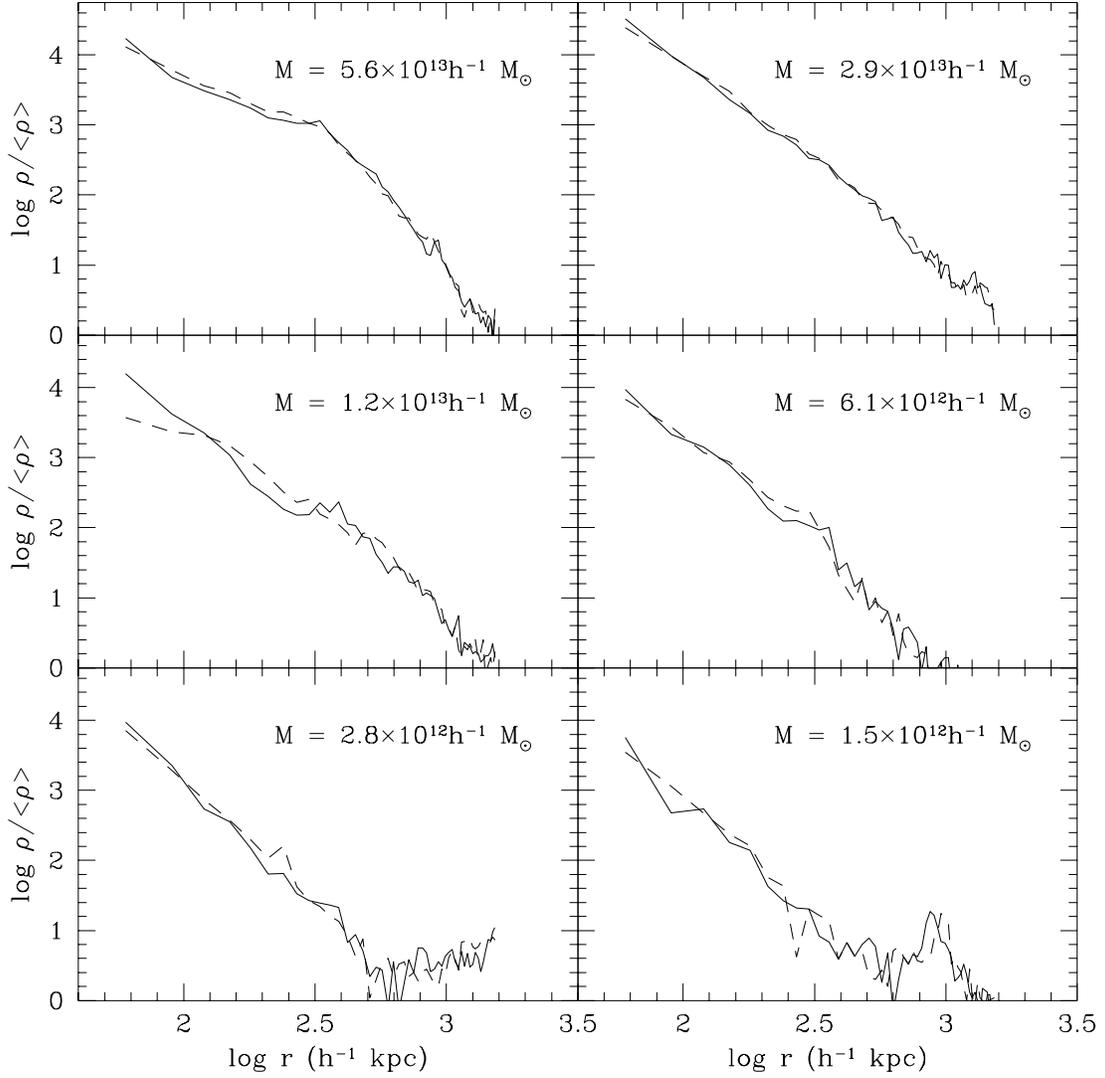}
\caption{Comparison of density profiles for halos identified 
in PM ({\em dashed lines}) and ART ({\em solid lines}) simulations 
($32^3$ particles) of comparable 
resolution ($\sim 60h^{-1}$ kpc) with the same initial 
conditions. 
} 
\label{pm2amr}
\end{figure}
\begin{figure}
\plotone{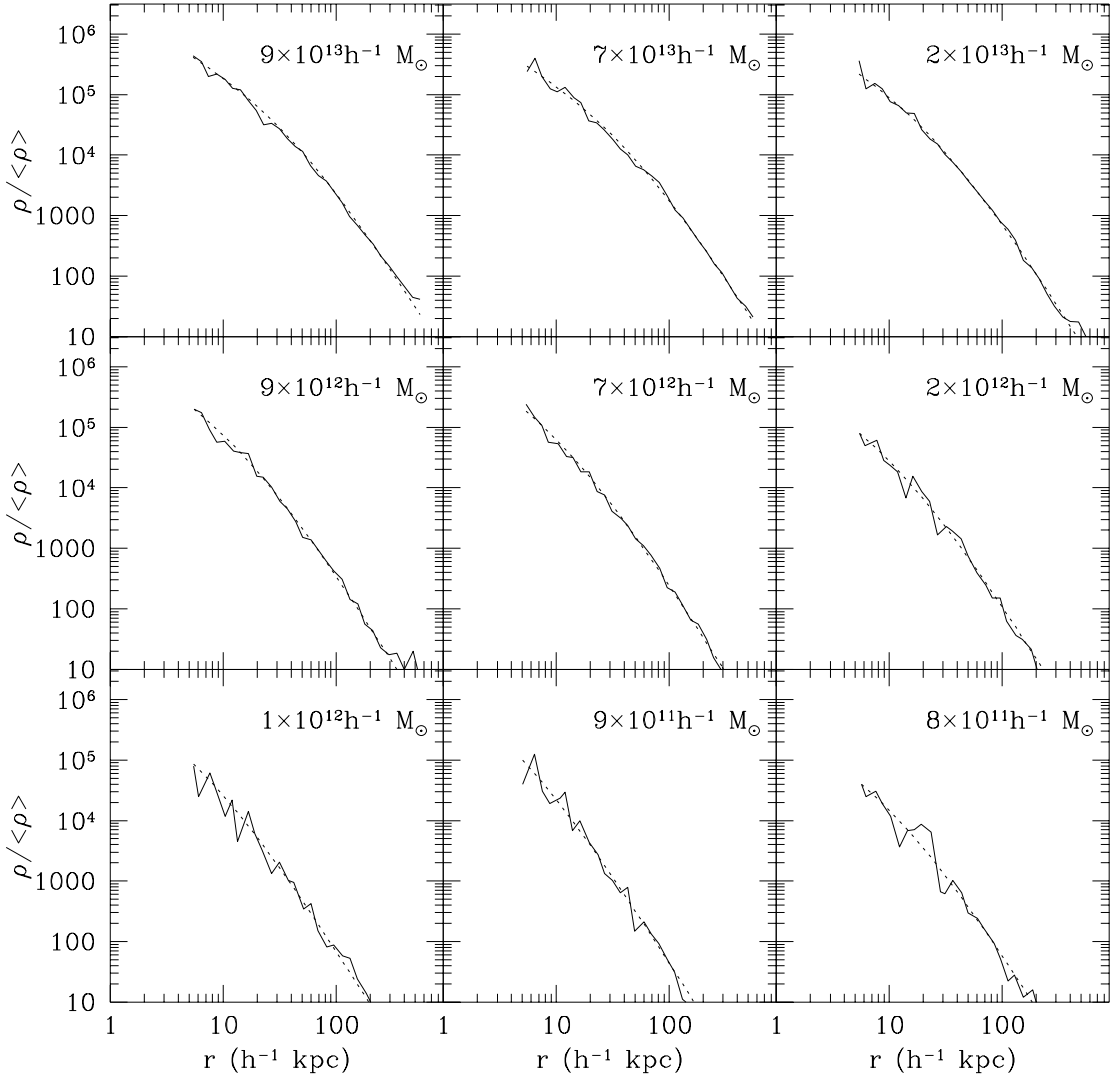}
\caption{{\em Solid lines}: Density profiles for nine halos of
different masses taken from the CDM  
simulation with $64^3$ particles (resolution $\sim 2h^{-1}$ kpc).
{\em Dotted lines}: Best fit by the analytic profile 
of Navarro et al. (1996a). Numbers in each panel indicate 
the mass of the halo inside
a radius corresponding to an overdensity of $200$. 
} 
\label{cdm_prof}
\end{figure}
\begin{figure}
\plotone{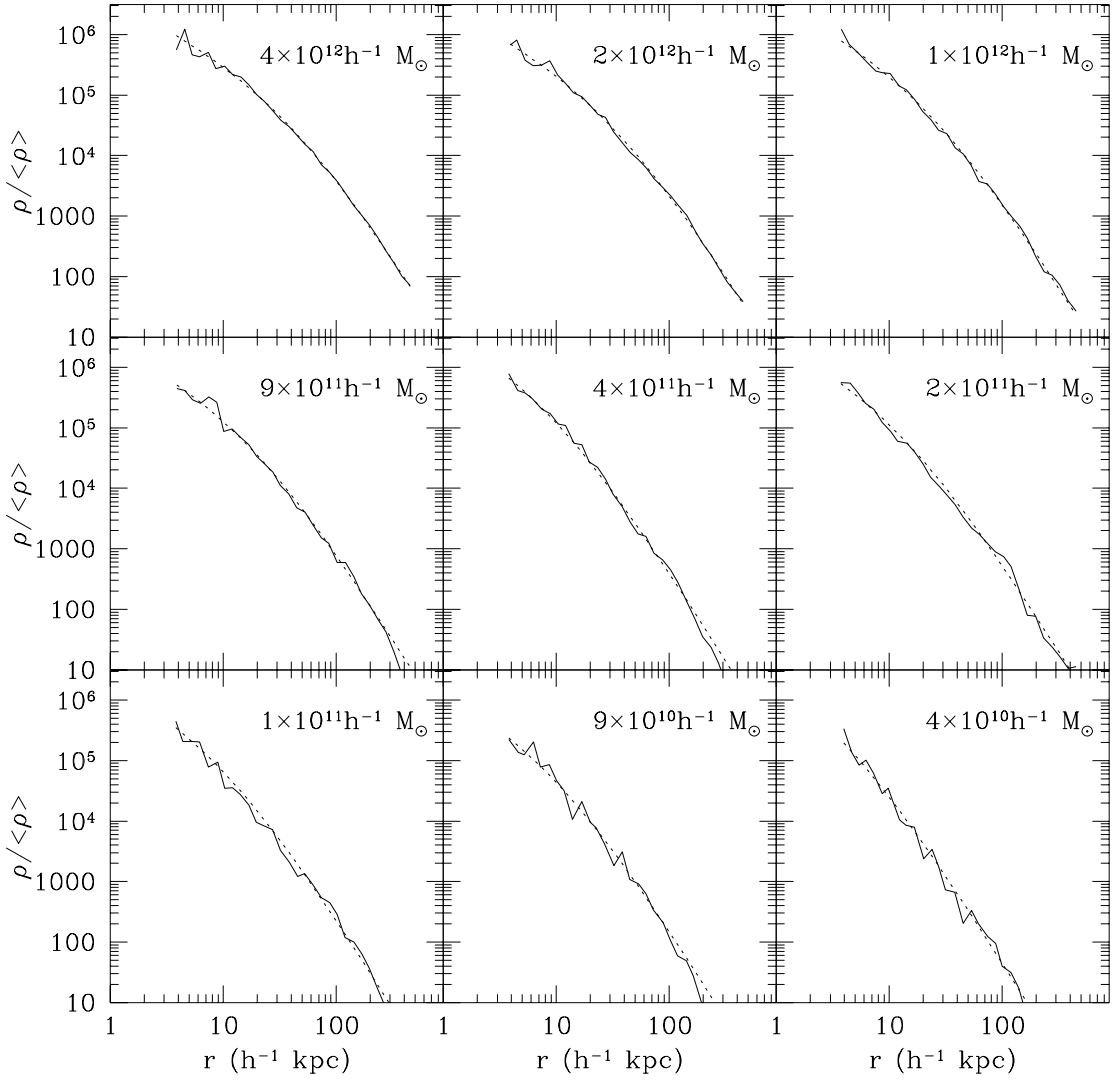}
\caption{Same as Fig.14, but for $\Lambda$CDM.
Numbers correspond to mass inside 
a radius of overdensity $334$ (see \S 5.4). 
} 
\label{lcdm_prof}
\end{figure}
\begin{figure}
\plotone{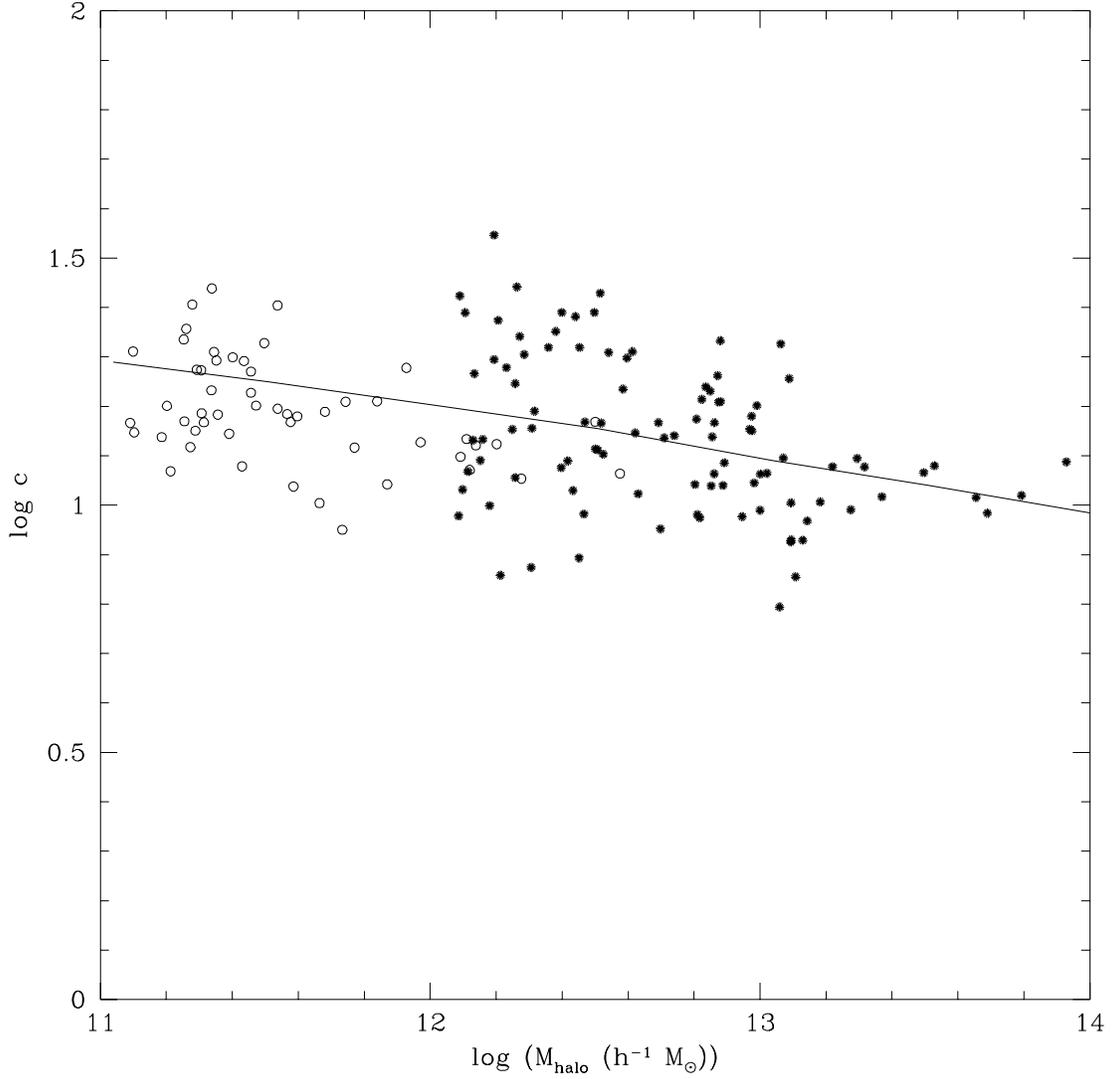}
\caption{ Logarithm of the concentration parameter, $c=r_{200}/r_s$,
vs. logarithm
of halo mass $M_{200}$ for high-resolution ($\sim 2h^{-1}$ kpc) 
CDM ({\em filled circles}) and $\Lambda$CDM simulations
({\em empty circles}). The solid curve shows the mass-concentration relation 
predicted from the formation times of halos that best fitted
the numerical
results of Navarro et al. (1996a). 
} 
\label{cm}
\end{figure}
\begin{figure}
\plotone{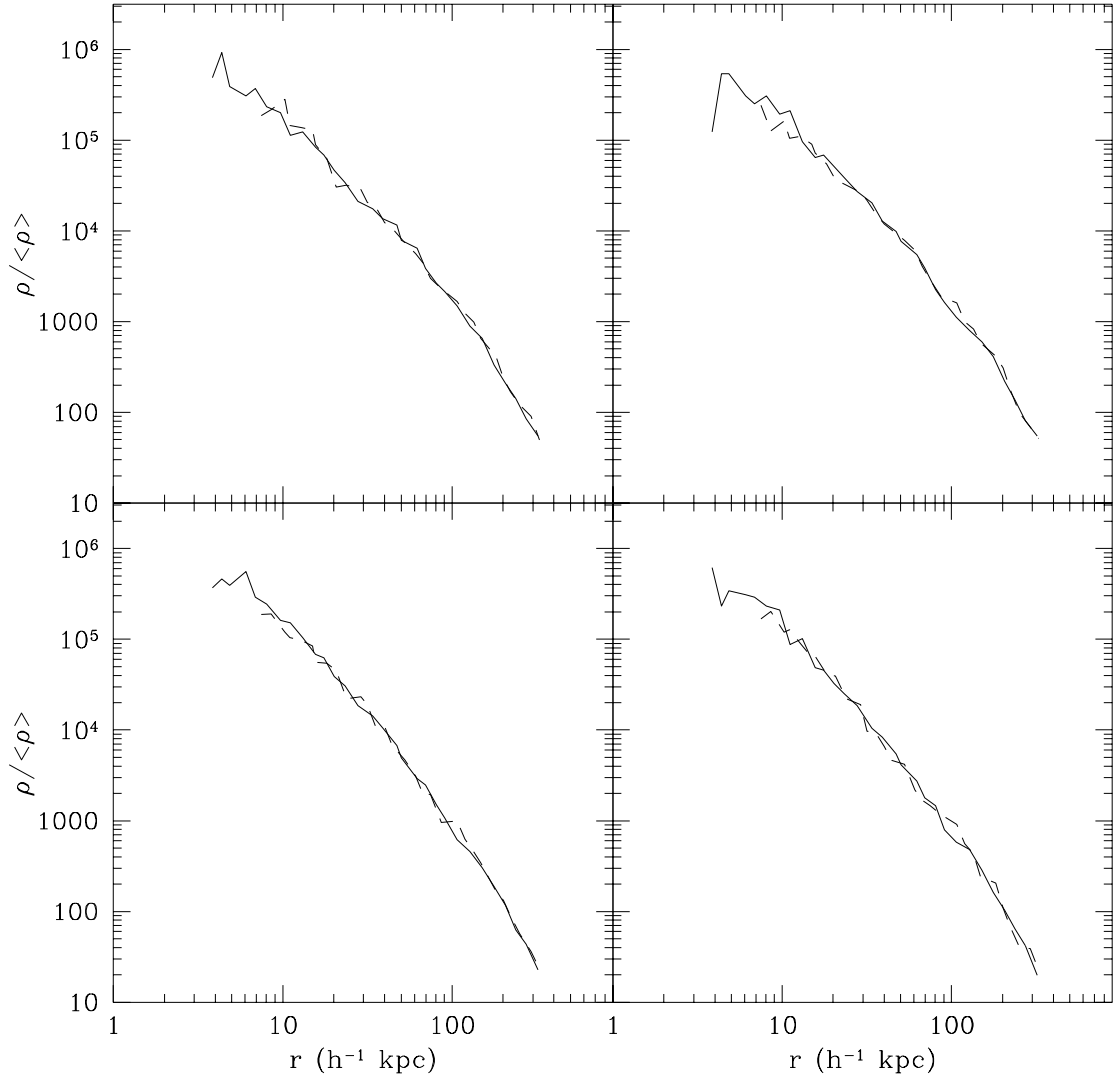}
\caption{Effects of resolution on halo density profiles. The profiles
of four halos taken from a $\Lambda$CDM run with $\sim 2h^{-1}$ kpc resolution 
({\em solid lines}) and a run with $\sim 7h^{-1}$ kpc resolution 
({\em dashed lines}) are
plotted down to the resolution limit. 
} 
\label{res}
\end{figure}
\clearpage
\begin{deluxetable}{ccccccccc}
\scriptsize
 \tablecaption{Code timings for $\Lambda$CDM simulations with 
  $64^3$ particles on 8 CPUs of the SPP-1200 Exemplar.}
\tablehead{
\colhead{Routine} & \colhead{Maximum} & 
\colhead{Number} & \colhead{Density} & 
\colhead{FFT} & \colhead{Relaxation} & 
\colhead{Particle} & \colhead{Mesh}& \colhead{Total}\nl
\colhead{ } & \colhead{level} & \colhead{of mesh cells} &
\colhead{assignment} & \colhead{solver} & \colhead{solver} & 
\colhead{motion} & \colhead{modifications} & 
}
\startdata
   RUN1 CPU (sec)    &   4      & 4860976    &  25.8     & 26.2 &  54.5      &  34.8    & 44.0 & 185.3 \nl
   RUN2 CPU (sec)    &   6      & 5019136    &  30.6     & 26.4 &  82.7      &  40.8    & 55.7 & 236.2 \nl
\enddata
\end{deluxetable}


\newpage
\begin{thebibliography}{}
\bibitem[Aho, Hopcroft, and Ullman 1983]{Aho}
             Aho, A.V., Hopcroft, J.E., \& Ullman, J.D. 1983,
             {\em Data Structures and Algorithms} (Reading: Addison-Wesley)
\bibitem[Anninos, Norman, and Clarke 1994]{ANC94}
             Anninos, P., Norman, M.L., \& Clarke, D.A. 1994,
             \apj {\ } 436, 11
\bibitem[Berger 1986]{Berger86} Berger, M.J.
             1986, SIAM J.Sci.Stat.Comput. 7, 904
\bibitem[Berger and Colella 1989]{BergerColella89}
             Berger, M.J., \& Colella, P. 1989,
             J.Comp.Phys. 82, 64
\bibitem[Berger and Oliger 1984]{BergerOliger84}
             Berger, M.J., \& Oliger, J. 1984,
             J.Comp.Phys. 53, 484
\bibitem[Bertschinger 1985]{bert85} Bertschinger, E. 
             1985, \apjs {\ } 58, 39
\bibitem[Bouchet and Hernquist 1988]{BH88} 
             Bouchet, F.R., \& Hernquist, L. 1988,
             \apjs {\ } 68, 521
\bibitem[Brandt 1977]{Brandt77} Brandt, A. 1977,
             Math. Comput., 31, 333
\bibitem[Burkert 1996]{Burkert96}
             Burkert, A., 1996, \apjl, 447, L25
\bibitem[Carrol et al. 1992]{Carroletal92}
             Carrol, S.M., Press, W.H., \& Turner, E.L. 1992,
             ARA\&A, 30, 499
\bibitem[Cole and Lacey 1996]{CL96}
             Cole, S., \& Lacey, C. 1996,
             \mnras, 281, 716
\bibitem[Corner, Leiserson, and Rivest 1994]{Corner}
             Corner, T.H., Leiserson, C.E., \& Rivest, R.L.
             1994, {\em Introduction to Algorithms} (New York: McGraw-Hill)
\bibitem[Couchman 1991]{Couchman91} Couchman, H.M.P. 1991,
             \apjl, 368, L23
\bibitem[Crone, Evrard, and Richstone 1994]{CER94}
             Crone, M.M., Evrard, A.E., \& Richstone, D.O., 1994,
             \apj \ 434, 402             
\bibitem[Dubinski \& Carlberg 1991]{DC91}
             Dubinski, J., \& Carlberg, R.G. 1991,
             \apj, 378, 496
\bibitem[Efstathiou et al.\ 1985]{EDFW85}
             Efstathiou, G., Davis, M., Frenk, C.S., \& White, S.D.M. 1985,
            \apjs, 57, 241
\bibitem[Efstathiou et al.\ 1988]{EFWD88}
             Efstathiou, G.,  Frenk, C.S., White, S.D.M., \& Davis, M. 1988,
            \mnras, 235, 715
\bibitem[Fillmore and Goldreich 1984]{FG84} Fillmore, J.A., \& Goldreich, P.
               1984, \apj, 281, 1
\bibitem[Flores and Primack 1994]{FP94}
             Flores, R.A., \& Primack, J.R. 1994,
             \apjl, 427, L1
\bibitem[Frenk et al. 1988]{Frenketal88}
            Frenk, C.S., White, S.D.M., Efstathiou, G., \& Davis, M., 
               1985, Nature, 317, 595
\bibitem[Frenk et al. 1988]{Frenketal88}
            Frenk, C.S., White, S.D.M., Davis, M., \& Efstathiou, G.
               1988, \apj, 327, 507
\bibitem[Gelato, Chernoff, and Wasserman 1996]{GCW}
             Gelato, S., Chernoff, D.F., \& Wasserman, I. 1996,
            \apj, 480, 115
\bibitem[Gelb92]{Gelb92} Gelb, J.M., 1992, PhD Thesis, MIT
\bibitem[Gnedin 1995]{Gnedin95} Gnedin, N.Y. 1995,
            \apjs, 97, 231
\bibitem[Gnedin and Bertschinger 1996]{GB96} 
            Gnedin, N.Y., \& Bertschinger, E. 1996,
            \apj, 470, 115
\bibitem[Hockney and Eastwood 1981]{HE} Hockney, R.W., \& Eastwood,J.W. 1981,
            {\em Computer simulations using particles} 
            (New York: McGraw-Hill)
\bibitem[Hoffman 1988]{Hoffman88}
            Hoffman, Y. 1988, \apj, 328, 489
\bibitem[Hoffman \& Shaham 1985]{HS85}
            Hoffman, Y., \& Shaham, J. 1985,
            \apj, 297, 16
\bibitem[Jessop, Duncan, and Chau 1994]{JDC}
            Jessop, C., Duncan, M., \& Chau, W.Y. 1994, 
            J.Comput.Phys., 115, 339
\bibitem[Kates, Kotok, and Klypin 1991]{KKK91} Kates, R.E., Kotok, E.V.,
            \& Klypin, A.A. 1991, \aap, 243, 295
\bibitem[Katz 1991]{Katz91}
            Katz, N., 1991, \apj, 368, 325
\bibitem[Kitayama \& Suto 1996]{KS96}
            Kitayama, T., \& Suto, Y. 1996,
            \apj, 469, 480
\bibitem[Khokhlov 1997]{Alexei} Khokhlov, A.M., 1997,
            J.Comput.Phys., submitted (preprint {\tt astro-ph/9701194})
\bibitem[Klypin 1996]{LectureI}
            Klypin, A.A. 1996, in
            International School of Physics 
            ``Enrico Fermi'': Dark Matter in the Universe, 
            ed. S.Bonometto, J.R.Primack,
            \& A.Provenzale, preprint {\tt astro-ph/9605183}
\bibitem[Klypin, Primack, and Holtzman 1996]{KPH96}
           Klypin, A.A., Primack, J.R., \& Holtzman, J. 1996, 
            \apj, 466, 13
\bibitem[Klypin and Shandarin 1983]{KS83} 
            Klypin, A.A., \& Shandarin, S.F. 1983, 
            \mnras, 204, 891
\bibitem[Knuth 1968]{Knuth1} Knuth, D. 1968, 
            {\em The Art of Computer Programming}, Vol.1 
            (Reading: Addison-Wesley)
\bibitem[Lahav et al. 1991]{Lahav91}
            Lahav, O., Lilje, P.B., Primack, J.R., \& Rees, M.J. 1991
       MNRAS, 251, 128L
\bibitem[L\"ohner and Baum 1991]{LohnerBaum91} 
            L\"ohner, R., \& Baum, J.D. 1991, AIAA J., 91-0620
\bibitem[Moore 1994]{Moore94}
            Moore, B. 1994, Nature, 370, 629
\bibitem[Navarro 1996]{Navarro96} 
            Navarro, J.F. 1996, preprint {\tt astro-ph/9610188}
\bibitem[Navarro, Frenk, and White 1996a]{NFW96}
            Navarro, J.F., Frenk, C.S., \& White, S.D.M., 1996a,
            \apj, 462, 563 (NFW)
\bibitem[Navarro, Frenk, and White 1996]{NFW96b}
            Navarro, J.F., Frenk, C.S., \& White, S.D.M., 1996b,
            ApJ, in press (preprint {\tt astro-ph/9611107})
\bibitem[Peebles 1980]{LSS}
            Peebles, P.J.E. 1980, 
            {\em The Large Scale Structure of the Universe}
            (Princeton: Princeton Univ. Press)
\bibitem[Pen 1995]{Pen95} Pen, U.-L. 1995,
            \apjs, 100, 269
\bibitem[Press et al.\ 1992]{NR92}
             Press, W.H., Teukolsky, S.A., Vetterling, W.T., \& Flannery, B.P.
             1992, {\em Numerical Recipes in FORTRAN}, 2nd ed.
\bibitem[Quinn, Salmon, and Zurek 1986]{QZS86}
             Quinn, P.J., Salmon, J.K., \& Zurek, W.H., 1986,
             Nature, 322, 329
\bibitem[Splinter 1996]{Splinter96} Splinter, R.J.
               1996, \mnras, 281, 281
\bibitem[Suisalu and Saar 1995]{SS95} Suisalu, I., \& Saar, E. 1995,
            \mnras, 274, 287
\bibitem[Suisalu and Saar 1996]{SS96} Suisalu, I., \& Saar, E. 1997,
            \mnras, submitted (preprint {\tt astro-ph/9511120})
\bibitem[Tormen, Bouchet, and White 1996]{TBW96}
       Tormen, G., Bouchet, F.R., White, S.D.M. 1996,
            \mnras, 286, 865
\bibitem[Villumsen 1989]{Vill89} Villumsen, J.V. 1989,
            \apjs, 71, 407
\bibitem[Warren et al. 1992]{Warrenetal92}
            Warren, M.S., Quinn, P.J., Salmon, J.K., \& Zurek, W.H. 1992,
            \apj, 399, 405
\bibitem[Wesseling 1992]{W92} Wesseling, P. 1992,
            {\em An Introduction to Multigrid Methods} (New York: Wiley)
\bibitem[Xu 1995]{Xu95} Xu, G. 1995,
            \apjs, 98, 355
\bibitem[Zel'dovich 1970]{Zeldovich70} Zeldovich, Ya.B., 1970,
            \aap, 5, 84

\end{thebibliography}
\end{document}